\documentclass[12pt]{article}

\usepackage{epsf}
\usepackage{latexsym,euscript}
\usepackage[dvips]{graphicx}
\usepackage{amsmath}
\usepackage{amsfonts}
\usepackage{amssymb, epsfig}
\usepackage{cite}
\usepackage{subcaption}
\usepackage[]{units}
\usepackage{mwe}
\usepackage{multirow}
\usepackage{verbatim}
\usepackage{geometry}
\usepackage{color,soul}

\usepackage{longtable}
\usepackage{booktabs}
\usepackage{float}
\usepackage{lscape}
\usepackage{subcaption}
\usepackage{cleveref}

\DeclareMathOperator{\arcsinh}{arcsinh}
\DeclareMathOperator{\Tr}{Tr}

\renewcommand{\Re}{\operatorname{Re}}
\renewcommand{\Im}{\operatorname{Im}}

\begin{document}

\begin{center}
  {\Large \bf Dual simulation of a Polyakov loop model at finite baryon
    density: \\ \vspace{0.1cm} phase diagram and local observables}
\end{center}

\vskip 0.3cm
\centerline{O.~Borisenko$^{1\dagger}$, V.~Chelnokov$^{1,2,3*}$,
E.~Mendicelli$^{4\ddagger}$, A.~Papa$^{2,5\P}$}

\vskip 0.6cm

\centerline{${}^1$ \sl Bogolyubov Institute for Theoretical Physics,}
\centerline{\sl National Academy of Sciences of Ukraine,}
\centerline{\sl 03143 Kyiv, Ukraine}

\vskip 0.2cm

\centerline{${}^2$ \sl Istituto Nazionale di Fisica Nucleare,
Gruppo collegato di Cosenza,}
\centerline{\sl I-87036 Arcavacata di Rende, Cosenza, Italy}

\vskip 0.2cm

\centerline{${}^3$ \sl Institut f\"ur Theoretische Physik, Goethe-Universit\"at
  Frankfurt,}
\centerline{\sl Max-von-Laue-Str. 1, 60438 Frankfurt am Main, Germany}

\vskip 0.2cm

\centerline{${}^4$ \sl Department of Physics and Astronomy, York University,}
\centerline{\sl Toronto, ON, M3J 1P3, Canada}

\vskip 0.2cm

\centerline{${}^5$ \sl Dipartimento di Fisica, Universit\`a della
Calabria,}
\centerline{\sl I-87036 Arcavacata di Rende, Cosenza, Italy}

\vskip 0.6cm

\begin{abstract} 
  Many Polyakov loop models can be written in a dual formulation which is free
  of sign problem even when a non-vanishing baryon chemical potential is
  introduced in the action. 
  Here, results of numerical simulations of a dual representation of one such
  effective Polyakov loop model at finite baryon density are presented. We
  compute various local observables such as energy density, baryon density,
  quark condensate and describe in details the phase diagram of the model. 
  The regions of the first order phase transition and the crossover, as well as
  the line of the second order phase transition, are established. 
  We also compute several correlation functions of the Polyakov loops.
\end{abstract}

\vfill
\hrule
\vspace{0.3cm}
{\it e-mail addresses}:
$^\dagger$oleg@bitp.kiev.ua, \  $^*$Volodymyr.Chelnokov@lnf.infn.it, \\
$^{\ddagger}$emanuelemendicelli@hotmail.it, \ \ $^{\P}$alessandro.papa@fis.unical.it

\newpage

\section{Introduction}

The properties of strongly interacting matter at finite temperatures and
densities remain in the focus of intensive theoretical and numerical studies
(see Ref.~\cite{philipsen19} for a recent review).  
The full understanding of these properties is still far from satisfactory,
especially at finite baryon chemical potential, due to the famous sign problem.
Many approaches to solve this problem, partially or completely, have been 
designed during the last decades and a certain progress has been achieved
within such methods as Taylor expansion and reweighting at small baryon
chemical potential, simulations at imaginary potential, complex Langevin
simulations and some others (see, {\it e.g.}, the
reviews~\cite{Schmidt:2006us,Forcrand_rev_10,Seiler_rev_17}).

One of the approaches attempting to fully solve the sign problem relies on
rewriting the original partition function and important observables in terms of
different, usually integer valued, degrees of freedom such that the resulting 
Boltzmann weight is positive definite. Conventionally, all such formulations
are referred to as dual formulations, though sometimes a different name can be
used ({\it e.g.}, flux line representation)~\cite{Gattringer_rev_16}. 
There are several routes to construct such dual theory for non-Abelian lattice
models with fermions~\cite{su3_abc,un_dual,Unger_19}. 
While a dual formulation with a positive Boltzmann weight has not yet
been constructed for full QCD, positive formulations (or formulations
where sign problem appears to be very soft) are already known for few
important cases. 
One such case refers to the strong coupling limit of QCD, where the $SU(N)$
lattice gauge theory can be mapped onto a monomer-dimer and closed baryon loop
model~\cite{Karsch_89} (for a recent development of this direction, see
Ref.~\cite{Unger_20} and references therein). Another important case is
represented by many effective Polyakov loop models which can be derived 
from the full lattice QCD in certain limits. A dual representation with
positive Boltzmann weight is known for some $SU(N)$~\cite{Gattringer11,Gattringer12,Philipsen12} and $U(N)$ Polyakov loop models~\cite{un_dual} 
(for recent advances, see Ref.\cite{Borisenko20}). Two of these versions have
been studied numerically in~\cite{Gattringer12,Philipsen12,Delgado12}. The
emphasis in these simulations was put on establishing the phase diagram 
of the model in the presence of the baryon chemical potential and on computing
local observables which can be obtained by differentiating the dual partition
function with respect to some of the parameters entering the action
of the theory.

Important class of observables not yet computed in dual formulations are
the correlations of the Polyakov loops. These correlations can be related to
screening (electric and magnetic) masses at finite temperatures. 
Understanding the properties of such masses would lead to an essential progress
in our comprehension of the high temperature QCD phase as a whole. While
correlations and related masses have been subject of numerous and intensive
calculations at zero chemical potential (see~\cite{Bazavov:2020teh} and
references therein), it seems to be an extremely difficult problem to compute
these masses for the real baryon chemical potential with available simulation
methods. So far correlations and screening masses have been computed only at
imaginary chemical potential in~\cite{bonati2018}. A closely related and
intriguing problem is the appearance of a hypothetical oscillating phase at
finite density~\cite{Ogilvie10,Ogilvie16,oscillating_phase}. Such a phase is
ultimately connected to the complex spectrum of the theory and requires
computations of long-distance correlations with real baryon chemical potential. 

Here and in a forthcoming paper we study a somewhat different, but equivalent
dual form of the effective Polyakov loop model presented in~\cite{Borisenko20}.
This form of the dual representation has been already used by us
in~\cite{Borisenko19} for the computation of correlation functions related to
the three-quark potential. We believe it is well suited to address the problem
of screening masses at finite densities, at least in the framework of the
available positive dual formulations. 
In the present paper we describe the Polyakov loop model we work with, its dual
representation and several observables. 
We compute also some local observables and reveal the phase structure of the
model. We shall also present preliminary results for the Polyakov loop
correlations. In a companion paper we will give a detailed study of screening
masses, based on the computation of correlation functions and of the second
moment correlation length at finite density. This will also allow us to
draw some conclusions about the existence of an oscillating phase in the
model. 

We work on a $3$-dimensional hypercubic lattice $\Lambda = L^3$,
with $L$ the linear extension and a unit lattice spacing.
The sites of the lattice are denoted by $\vec{x}\equiv x=(x_1,x_2,x_3)$,
$x_i\in [0,L-1]$, while $l=(\vec{x},\nu)$ is the lattice link in the
direction $\nu$; $e_{\nu}$ is the unit vector in the direction $\nu$ and
$N_t$ is the lattice size in the temporal direction.
Periodic boundary conditions are imposed in all directions. 
Let $G$ be the $SU(N)$ group and $U(x)$ an element of $G$, then $dU$ denotes
the (reduced) Haar measure on $G$ and ${\rm Tr}U$ the fundamental character
of $G$. 

In this paper we shall study an effective 3-dimensional Polyakov loop model
which describes a $(3+1)$-dimensional lattice gauge theory with one flavor of
staggered fermions.
The general form of the partition function of the model is given by  
\begin{eqnarray} 
\label{PF_spindef}
Z_{\Lambda}(\beta,m,\mu;N,N_f)  \equiv  Z \ = \ 
\int \ \prod_x \ dU(x) \prod_{x,\nu} \ B_{\rm g}(\beta) \ 
\prod_x \ B_{\rm q}(m,\mu) \ . 
\end{eqnarray}
Here, $B_{\rm g}(\beta)$ is the gauge part of the Boltzmann weight and
$B_{\rm q}(m,\mu)$ is the determinant for static quarks. 
There are many forms of $B_{\rm g}(\beta)$ and $B_{\rm q}(m,\mu)$ discussed in the
literature. In what follows we use the weight that can be obtained on an
anisotropic lattice and in the limit of vanishing spatial gauge coupling
$\beta_{\rm s}$ after explicit integration over all spatial gauge fields
(see, for instance, Refs.~\cite{Caselle97,Philipsen12,Philipsen14} and
references therein):
\begin{equation}
  B_{\rm g}(\beta) \ = \  \exp \left [ \beta \ {\rm Re}{\rm Tr}U(x)
    {\rm Tr}U^{\dagger}(x+e_{\nu}) \right ] \ . 
\label{Bgauge_strcpl}
\end{equation}
For $SU(N)$ the effective coupling constant $\beta$ is related to the temporal
coupling $\beta_{\rm t}$ by $\beta=2D_{\rm F}(\beta_{\rm t})$ with  
\begin{eqnarray}
  D_{\rm F}(\beta_{\rm t}) \ = \ \left ( \frac{C_{\rm F}(\beta_{\rm t})}{N C_{0}(\beta_{\rm t})}
  \right )^{N_t}
  \ , \ \ \ 
  C_{\rm F}(\beta_{\rm t}) \ = \ \sum_{k=-\infty}^{\infty}
  \ \Bigl.{\rm det} I_{\lambda_i - i + j + k}(\beta_{\rm t})\Bigr|_{1\leq i,j \leq N} \ , 
\label{D_coeff}
\end{eqnarray}
where $I_n(x)$ is the modified Bessel function and $\lambda_i$ refers to
the fundamental representation of $SU(N)$ and is equal to
$\lambda_i=\delta_{1i}$.
The Boltzmann weight of static staggered fermions can be presented as 
\begin{equation}
B_{\rm q}(m,\mu) \ = \ 
A(m) \ {\rm det} \left [ 1 + h_+ U(x) \right ] \ {\rm det} \left [ 1 + h_- U^{\dagger}(x) \right ] \ ,
\label{Zf_stag_massive}
\end{equation}
where the determinant is taken over group indices and 
\begin{equation}
A(m) = h^{-N} \ , \ \ \ h_{\pm} = h e^{\pm \frac{\mu_{ph}}{T}} \ , \ \ \ 
h = e^{-N_t \arcsinh m} \approx e^{-\frac{m_{ph}}{T}} \ . 
\label{hpm_stag}
\end{equation}
Below we shall use the dimensionless quantities $m=m_{\rm ph}/T$ and
$\mu=\mu_{\rm ph}/T$. Then for $SU(3)$ one has $A(m)=e^{3 m}$. 

The resulting Polyakov loop model we work with takes the form 
\begin{eqnarray}
Z  \ &=& \ \int \prod_x dU(x)
\exp \left [ \beta \sum_{x,\nu} \ {\rm {Re}}{\rm {Tr}}U(x){\rm {Tr}}U^{\dagger}
  (x+e_\nu) \right ] \nonumber  \\ 
&\times& \ \prod_x \ A(m) \ {\rm det} \left [ 1 + h_+ U(x) \right ] \ {\rm det}
\left [ 1 + h_- U^{\dagger}(x) \right ] \ . 
\label{sunpf}
\end{eqnarray}
In this model the matrices $U(x)$ play the role of Polyakov loops, the only
gauge-invariant operators surviving the integration over spatial gauge fields
and over quarks. 
The integration in~(\ref{sunpf}) is performed with respect to the Haar measure
on $G$. 
The pure gauge part of the $SU(N)$ model is invariant under global discrete
transformations $U(x)\to Z U(x)$, with $Z\in Z(N)$. This is the global $Z(N)$
symmetry. The quark contribution violates this symmetry explicitly. 
Another important feature of the Boltzmann weight is that it becomes complex in
the presence of a chemical potential, as it follows from~(\ref{sunpf}).
Therefore, the model cannot be directly simulated if $\mu$ is non-zero. 

In the absence of static quarks the Polyakov loop model exhibits a first order
phase transition at the critical point $\beta_{\rm c}\approx 0.274$. The global
$Z(N)$ symmetry gets spontaneously broken above $\beta_{\rm c}$.
At finite density
the model defined in Eq.~(\ref{sunpf}) and its several variations have been
studied both numerically via simulation of dual
formulations~\cite{Gattringer12,Philipsen12,Delgado12} and analytically via
mean-field approximation~\cite{Greensite12,Greensite14,Rindlisbacher:2015pea}
and via linked cluster expansion~\cite{Kim:2020atu}.
Mean-field and Monte Carlo study are in quantitative agreement for the
expectation values of energy density and Polyakov loop. 
This allowed to reveal the phase diagram of the model, at least in some regions
of the parameters $\beta$, $h$ and $\mu$. 
In this paper we confirm the qualitative picture found in previous study and
give further details on the behavior of local observables, including the baryon
density and the quark condensate. Also, first results for Polyakov loop
correlations will be presented. 

The paper is organized as follows.  
In Section~2 we formulate our dual representation of the model valid for all
$SU(N)$ groups and in all dimensions. 
We present also results of an analytical study of the model based on strong
coupling expansion and mean-field approximation. 
The phase diagram of the 3-dimensional $SU(3)$ model is studied numerically
in details in Section~3, where we discuss also simulation results for some local
observables as the baryon density and the quark condensate.
In Section~4 we present preliminary results for the Polyakov loop correlation
functions. The summary and outline for future work is done in Section~5. 

\section{Dual formulation of the Polyakov loop model}

In this Section we describe the dual form of the partition
function~(\ref{sunpf}). This dual representation will be used 
in the next Sections for numerical simulations of the model. All details of the
derivation can be found in~\cite{Borisenko20}. 
In the case of one flavor of staggered fermions the partition
function~(\ref{sunpf}) can be presented, after an exact integration over
Polyakov loops, as  
\begin{equation} 
\label{PF_statdet_3}
Z = \sum_{\{ r(l) \} = -\infty}^{\infty} \ \sum_{\{ s(l) \}= 0}^{\infty} \ 
\prod_l  \frac{\left ( \frac{\beta}{2} \right )^{| r(l) | + 2s(l)}}{(s(l)+| r(l) |)!
  s(l)!} \ 
\prod_x A(m) R_N(n(x),p(x))  \ , 
\end{equation}
\begin{eqnarray}
\label{nx_stat}
&&n(x) =  \sum_{i=1}^{2d}  \left ( s(l_i) + \frac{1}{2} | r(l_i) | \right )
+ \frac{1}{2} \sum_{\nu=1}^{d} \left ( r_{\nu}(x) - r_{\nu}(x-e_{\nu}) \right )
\ ,   \\ 
\label{px_stat}
&&p(x) =  \sum_{i=1}^{2d}  \left ( s(l_i) + \frac{1}{2} | r(l_i) | \right )
- \frac{1}{2} \sum_{\nu=1}^{d} \left ( r_{\nu}(x) - r_{\nu}(x-e_{\nu}) \right )
\ ,  
\end{eqnarray} 
where $l_i, i=1,...,2d$ are $2d$ links attached to a site $x$ and 
\begin{eqnarray} 
  R_N(n,p) = \sum_{q=-\infty}^{\infty} \sum_{k,l=0}^N \sum_{\sigma \vdash n+k}
  \ \delta_{n+k,p+l+qN} \ 
d(\sigma/1^k) d(\sigma+q^N/1^l) \ h_+^k h_-^l \ . 
\label{Rint_Nf1}
\end{eqnarray}
The sum over $\sigma$ runs over all partitions of $n+k$, and
$d\left ( \sigma/1^m  \right )$ is the dimension of a skew representation
defined by a corresponding skew Young diagram,
$\sigma+q^N=(\sigma_1+q,\ldots,\sigma_N+q)$
(for more details we refer the reader to Ref.~\cite{Borisenko20}).
Equation~({\ref{PF_statdet_3}}) is valid for all $SU(N)$ groups and in any
dimension. 
Clearly, all factors entering the Boltzmann weight of~(\ref{PF_statdet_3}) are
positive. 
Hence, this representation is suitable for numerical simulations.
The Kronecker delta-function in expression~(\ref{Rint_Nf1}) represents the
$N$-ality constraint on the admissible configurations of the integer-valued
variables $s(l)$ and $r(l)$. 
This constraint can be exactly resolved only in the pure gauge model when
$h_{\pm}=0$. 
In this case the dual representation~(\ref{PF_statdet_3}) has been already
tested by us on an example of 2-dimensional $SU(3)$ model, where we studied
correlation functions and three-quark potential~\cite{Borisenko19}. 

In the following Sections we study the dual representation~(\ref{PF_statdet_3})
via Monte Carlo simulations for the 3-dimensional $SU(3)$ model. In this
case the function $R_N(n,p;h_{\pm})$ takes the form 
\begin{eqnarray} 
\label{Rint_N3_Nf1}
&&R_3(n,p) = Q_3(n+1,p) \left ( h_+ + h_-^2 + h_+ h_-^3 + h_+^3 h_-^2  \right ) \\
&+&Q_3(n,p) \left ( 1+ h_+^3 + h_-^3 + h_+^3 h_-^3  \right )  + 
Q_3(n,p+1) \left ( h_- + h_+^2 + h_+^3 h_- + h_+^2 h_-^3  \right )   \nonumber \\ 
&+&Q_3(n+1,p+1) \left ( h_+ h_- + h_+^2 h_-^2  \right ) + 
Q_3(n+2,p) h_+ h_-^2  + Q_3(n,p+2) h_+^2 h_- \ .   \nonumber  
\end{eqnarray}
The function $Q_3(n,p)$ is the result of the group integration and is given
by~\cite{weingarten_sun}
\begin{equation} 
Q_N(n,p) \ = \  
\sum_{\lambda \vdash {\rm min}(n,p)} \ d(\lambda) \ d(\lambda + |q|^N) \ , 
\label{QSUN}
\end{equation}
where $d(\lambda)$ is the dimension of the permutation group $S_r$ in
the representation $\lambda$, $q = (p - n) / N$ (when $q$ is not an integer
$Q_N(n,p) = 0$).

Important is the fact that {\em both} local observables and long-distance
quantities can be computed with the help of this dual representation. Explicit
expressions for the correlation functions of the Polyakov loops 
will be given in Section~4. Below we list some local observables 
which we are going to compute here and which can be obtained by taking
suitable derivatives of the partition function~(\ref{PF_statdet_3}). 

\begin{itemize}
\item 
Magnetization and its conjugate
\begin{eqnarray}
\label{magnetization_dual}
M \ = \ \left\langle \Tr U(x) \right\rangle \ , \quad M^* \
= \ \left\langle \Tr U^\dagger (x) \right\rangle \ ,
\end{eqnarray}
\item 
Susceptibility
\begin{eqnarray}
\chi \ = \ 
L^2 \left( \left\langle \left(\Tr U(x)\right)^2 \right\rangle -
\left\langle \Tr U(x)\right\rangle^2 \right) \ ,
\label{susceptibility_dual}
\end{eqnarray}
\item 
Energy density
\begin{equation}
E \ = \ \frac{1}{3 L^3} \; \frac{\partial \ln Z}{\partial\beta} 
\ = \ \frac{2}{3 \beta L^3} \; \sum_l \langle 2 s(l) + |r(l)|  \rangle  \ ,
\label{energy_dual}
\end{equation}

\item 
Baryon density 
\begin{equation}
B \ = \ \frac{1}{L^3} \; \frac{\partial \ln Z}{\partial \mu} 
\ = \ \frac{1}{L^3} \; \sum_x \langle k(x) - l(x) \rangle \ ,
\label{baryonden_dual}
\end{equation}

\item 
Quark condensate 
\begin{equation}
Q \ = \ \frac{1}{L^3} \; \frac{\partial \ln Z}{\partial m} 
\ = \ N - \frac{1}{L^3} \; \sum_x \langle  k(x) + l(x) \rangle  \ .
\label{q_cond_dual}
\end{equation}

\end{itemize}
In the last two equations $k(x)$ and $l(x)$ are the summation variables
from~(\ref{Rint_Nf1}). 

Before discussing results of numerical simulations we would like to present
some results obtained by simple analytical methods. These results can serve as
an additional check of numerical data and lead to a better understanding of
the whole phase diagram of the model and of the behavior of the different 
expectation values. 

\subsection{Strong coupling expansion}

The formulation~(\ref{PF_statdet_3}) allows a straightforward expansion in
powers of $\beta$.
The free energy of the 3-dimensional $SU(3)$ model can be written as
\begin{equation}
 F = 3 m + \ln R_3(0,0) + \sum_{k=1}^{\infty} \beta^k f_k \ . 
\label{strong-coupling}
\end{equation}
The first three coefficients $f_k$ are given by
\begin{align}
f_1 &= 3 \, p_{0 1} \, p_{1 0} \nonumber \\
f_2 &= \frac{3}{4}
\left(
 p_{1 1}^2 + p_{0 2} \, p_{2 0} - 22 \, p_{0 1}^2 \, p_{1 0}^2
 + 5 \, p_{0 2} \, p_{1 0}^2 + 5 \, p_{0 1}^2 \, p_{2 0} + 10 \, p_{1 0}
\, p_{0 1} \, p_{1 1} \right) \nonumber \\
f_3 &= \frac{1}{8} \left( 1168 \, p_{0 1}^3 \, p_{1 0}^3 - 960 \, p_{0
1}^2 \, p_{1 0}^2 \, p_{1 1}
 - 480 \left(p_{0 1} \, p_{0 2} \, p_{1 0}^3 + p_{0 1}^3 \, p_{1 0} \,
p_{2 0} \right) \right. \nonumber \\
 &\quad \left. {} + 20 \left(p_{0 3} \, p_{1 0}^3 + p_{0 1}^3 \, p_{3
0}\right)
                  + 150 \, p_{1 1} \left( p_{0 2} \, p_{1 0}^2  + p_{0
1}^2 \, p_{2 0} \right) \right. \nonumber \\
 &\quad \left. {} + 84 \, p_{0 1} \, p_{1 0} \left( p_{1 1}^2 + p_{0 2} \,
p_{2 0} \right)
                  + 60 \left( p_{0 1} \, p_{1 0}^2 \, p_{1 2} + p_{0 1}^2
\, p_{1 0} \, p_{2 1} \right) \right. \nonumber \\
 &\quad \left. {} + 30 \, p_{1 1} \left(p_{1 0} \, p_{1 2} + p_{0 1} \,
p_{2 1} \right)
                  + 15 \, p_{2 0} \left(p_{0 3} \, p_{1 0} + p_{0 1} \,
p_{1 2} \right) \right. \nonumber \\
 &\quad \left. {} + 15 \, p_{0 2} \left(p_{0 1} \, p_{3 0} + p_{1 0} \,
p_{2 1} \right)
                  + p_{0 3} \, p_{3 0} + 3 p_{1 2} \, p_{2 1} \right)\ ,
\label{strong-coupling-coefficients}
\end{align}
where $p_{k l} = R_3(k,l) / R_3(0,0)$ and the coefficients $R_3$
can be easily calculated from Eqs.~(\ref{Rint_N3_Nf1}) and~(\ref{QSUN}),
resulting in
\begin{align}
R_3(0, 0) &= 1 + h^2 + h^4 + h^6 + 2 h^3 \cosh (3 \mu) \ , \nonumber \\
R_3(0, 1) &= h^2 e^{-2 \mu} \left(1 + h^2\right) + h e^{\mu} \left(1 + h^2
+ h^4\right) \ , \nonumber \\
R_3(1, 0) &= h^2 e^{2 \mu} \left(1 + h^2\right) + h e^{-\mu} \left(1 + h^2
+ h^4\right) \ ,\nonumber \\
R_3(0, 2) &= h e^{-\mu} \left(1 + h^2\right)^2 + h^2 e^{2 \mu} \left(1 +
h^2\right) \ , \nonumber \\
R_3(2, 0) &= h e^{\mu} \left(1 + h^2\right)^2 + h^2 e^{-2 \mu} \left(1 +
h^2\right) \ , \nonumber \\
R_3(1, 1) &= 1 + 2 h^2 + 2 h^4 + h^6 + 2 h^3 \cosh (3 \mu) \ , \nonumber \\
R_3(0, 3) &= R(3, 0) = \left(1 + h^2\right)^3 + 2 h^3 \cosh (3 \mu) \ ,
\nonumber \\
R_3(1, 2) &= 2 h^2 e^{-2 \mu} \left(1 + h^2\right) + h e^{\mu} \left(2 + 3
h^2 + 2 h^4\right) \ ,\nonumber \\
R_3(2, 1) &= 2 h^2 e^{2 \mu} \left(1 + h^2\right) + h e^{-\mu} \left(2 + 3
h^2 + 2 h^4\right) \ .
\label{small-R3}
\end{align}
All local observables listed above can be obtained from the
expansion~(\ref{strong-coupling}). 
The strong coupling expansion converges, presumably in the region
$\beta\leq\beta_{\rm c}(h,\mu)$, where  $\beta_{\rm c}(h,\mu)$ is the phase
transition or crossover point.
One expects that in this region numerical data agree reasonably well with
strong coupling results. For the purposes of this paper it
  was sufficient to consider only the lowest three orders in the strong
  coupling series. Higher orders can be calculated using the linked cluster
  expansion developed in~\cite{Kim:2020atu} for a similar Polyakov loop model.

\subsection{Mean-field solution} 

\begin{figure}[htb]
\centering
{
\hfill
\includegraphics[width=0.48\textwidth]{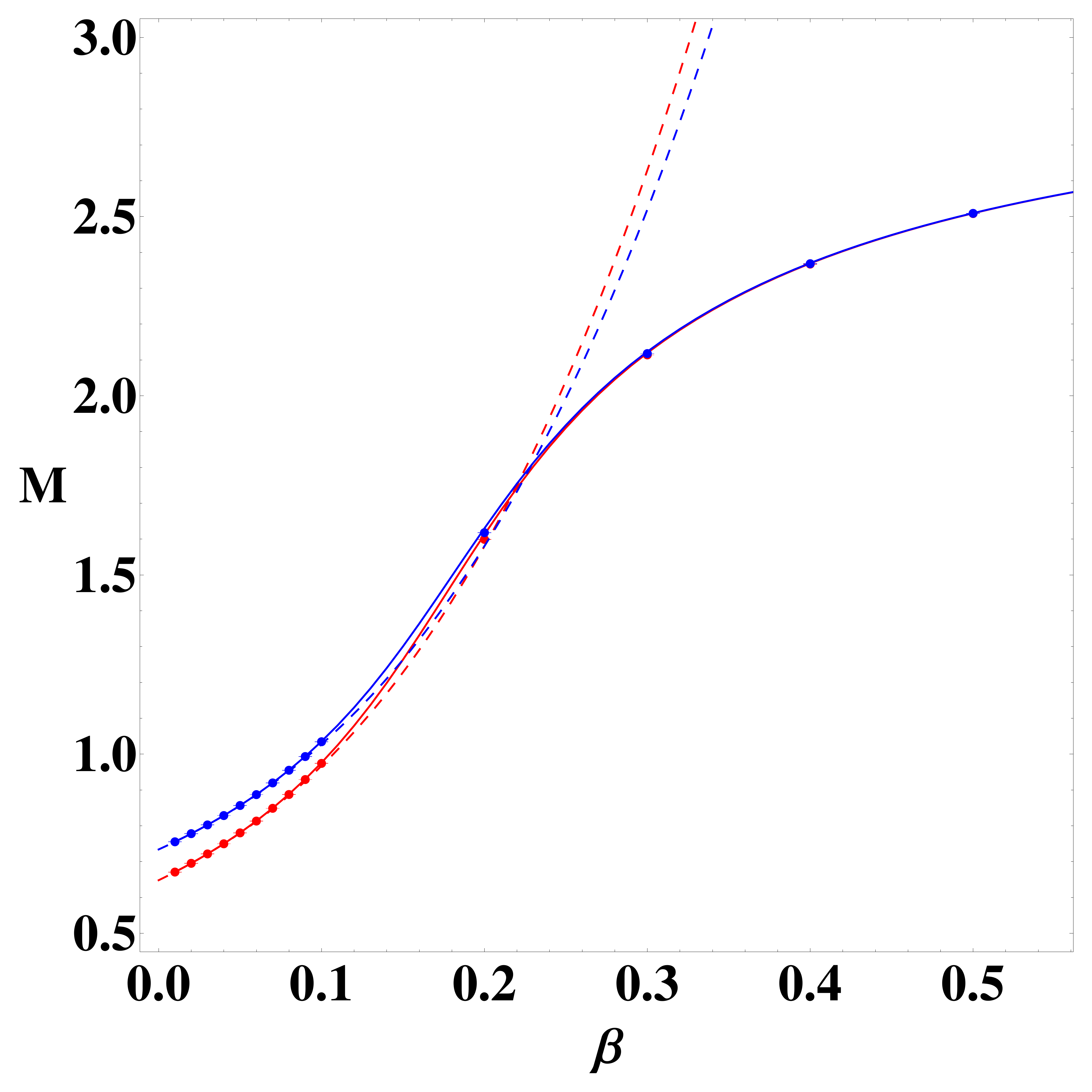}
\hfill
\includegraphics[width=0.48\textwidth]{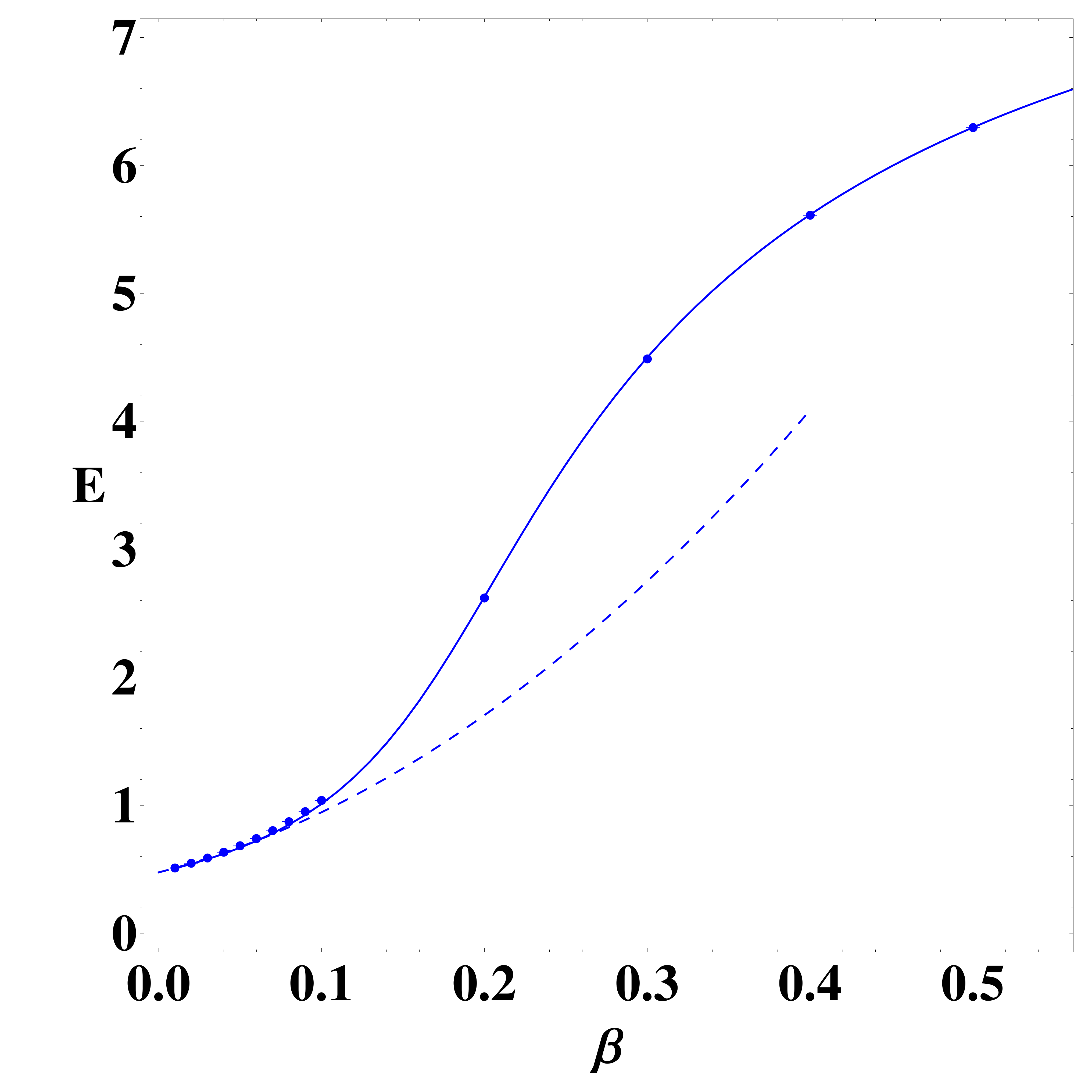}
\hfill \\
\hfill
\includegraphics[width=0.48\textwidth]{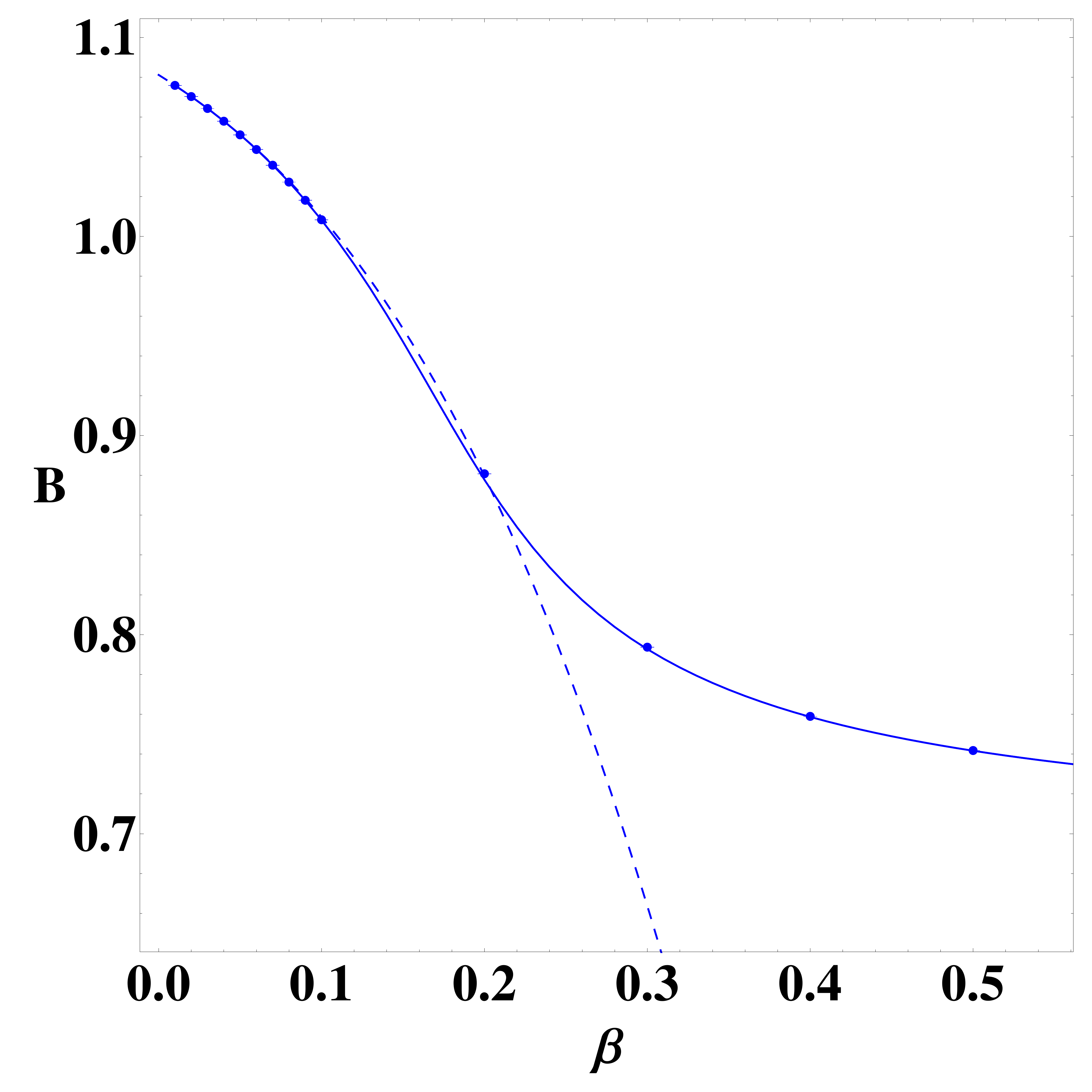}
\hfill
\includegraphics[width=0.48\textwidth]{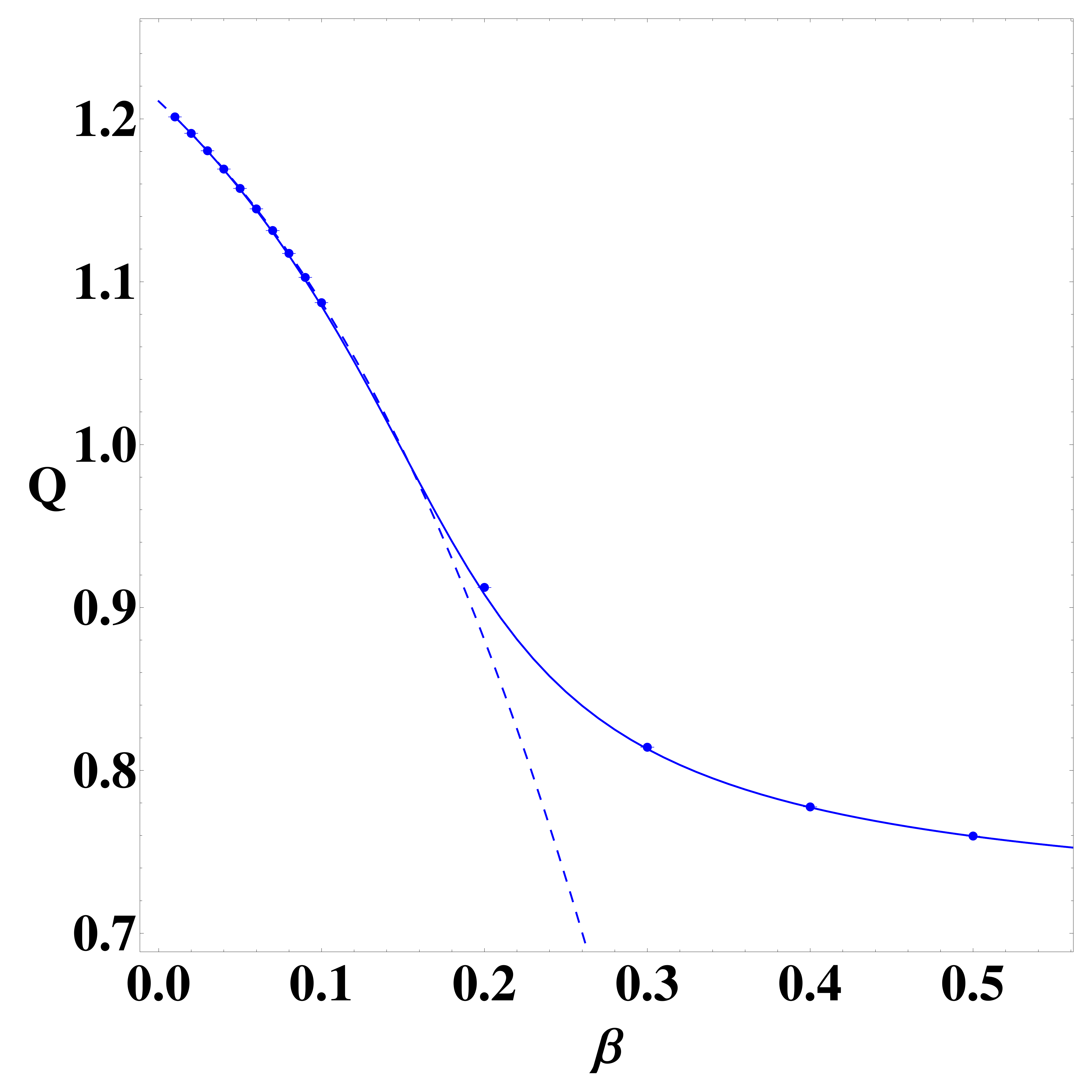}
\hfill 
}
\caption{Comparison of the observables obtained from mean-field analysis (solid
  lines), strong coupling expansion up to third order in $\beta$ (dashed lines)
  and simulation results (dots) at $h=0.6$, $\mu=0.5$ on a 16$^3$ lattice.
  Top left: magnetization (blue and red lines denote magnetization and its
  conjugate, respectively); top right: energy density; bottom left:
  baryon density; bottom right: quark condensate.}
\label{fig:comparison-high-h}
\end{figure}

Another obvious approach which can be used to get qualitative description of
the model~(\ref{sunpf}) is mean-field approximation. Within the mean-field
method one obtains an approximate phase diagram of the model. Also, it allows
to calculate various local observables, as the free energy density, the baryon
density and some others. We use here one of the simplest mean-field schemes,
applied to a similar Polyakov loop model
in~\cite{Greensite12,Greensite14,Rindlisbacher:2015pea }.
In this scheme the mean-field approximation reduces to the following
replacement: 
\begin{equation}
  \sum_{x,\nu} \ {\rm {Re}}{\rm {Tr}}U(x){\rm {Tr}}U^{\dagger}(x+e_\nu)
  \longrightarrow  
  \frac{2d}{2} \sum_{x} \left ( \omega {\rm {Tr}}U(x) + u {\rm {Tr}}U^{\dagger}(x)
  \right ) \ , 
\label{mean_field1}
\end{equation}
where 
\begin{equation}
  u =  \langle {\rm {Tr}}U(x) \rangle = \frac{1}{d\beta} \
  \frac{\partial \ln Z_{\rm mf}(u,\omega)}{\partial \omega} \ , \ 
  \omega = \langle {\rm {Tr}}U^{\dagger}(x) \rangle = \frac{1}{d\beta}
  \ \frac{\partial \ln Z_{\rm mf}(u,\omega)}{\partial u} \ . 
\label{mean-field_def}
\end{equation}
The partition function gets the form 
\begin{eqnarray}
Z  \ = \ \left [ Z_{\rm mf}(u,\omega)   \right ]^{L^d} \ , 
\label{sunpf_mf}
\end{eqnarray}
\begin{eqnarray}
Z_{\rm mf}(u,\omega)  \ &=& \ A(m) \ \int dU
\exp \left [ d \beta  \left ( \omega {\rm {Tr}}U + u {\rm {Tr}}U^{\dagger}
  \right )\right ] \nonumber  \\ 
&\times& \ {\rm det} \left [ 1 + h_+ U \right ] \ {\rm det}
\left [ 1 + h_- U^{\dagger} \right ] \ . 
\label{sunpf_mf_def}
\end{eqnarray}
Using the integration methods developed in~\cite{weingarten_sun,Borisenko20}
the mean-field partition function is presented as 
\begin{equation}
  Z_{\rm mf}(u,\omega) \ = \ \sum_{r,s=0}^{\infty} \ \frac{(d\beta\omega)^r}{r!}
  \ \frac{(d\beta u)^s}{s!}  \ R_3(r,s) \ 
\label{pf_mean-field}
\end{equation}
and explicit form of $R_3(r,s)$ reads 
\begin{eqnarray}
  R_3(r,s) = \sum_{i,j=0}^1 \ C_{ij} \ Q_3(r+i,s+j) + C_{20} \ Q_3(r+2,s) + C_{02}
  \ Q_3(r,s+2) \ , 
\label{R3_mf}
\end{eqnarray}
where the coefficients $C_{ij}$ are given by 
\begin{eqnarray}
C_{00} &=& 2 \cosh 3m + 2 \cosh 3\mu \ , 
C_{11}  = 2 \cosh m \ , \ 
C_{20} = e^{\mu} \ , \  C_{02} \ = \ e^{-\mu}  \ , \nonumber  \\ 
\label{Cij}
C_{10} &=& 2 e^{-\mu} \cosh 2m + 2 e^{2 \mu} \cosh m \ , 
C_{01} =  2 e^{\mu} \cosh 2m + 2 e^{-2 \mu} \cosh m \  . 
\end{eqnarray}
The mean-field equations~(\ref{mean-field_def}), as well as all local
observables, can now be computed numerically and compared with the strong
coupling results and numerical data. In the pure gauge case, $h=0$, one finds
a first order phase transition at the critical value $\beta_{\rm g}\approx
0.2615$.
This value, as well as the magnetization, matches the corresponding results
obtained earlier by the mean-field method~\cite{Greensite12,Greensite14}.
Fig.~\ref{fig:comparison-high-h} compares our numerical data with the strong
coupling expansion and mean-field results for some typical values
$h=0.6,\mu=0.5$. More mean-field results and comparison with 
simulations will be given in the next Section. 

\section{Phase diagram of the model} 

\subsection{Lattice setup}

To explore the phase structure of the model, we simulate numerically
the partition function~(\ref{PF_statdet_3}). Important ingredient of such 
simulations is the way we treat the triality constraint in~(\ref{Rint_Nf1}). 
In the pure gauge theory, $h=0$, this constraint is solved exactly in terms of 
genuine dual variables and the resulting theory can be simulated with 
the usual Metropolis update~\cite{Borisenko19}. Instead, at non-zero values of
$h$, the function $R_N(n,p)$ in~(\ref{PF_statdet_3}) is explicitly expanded 
in series~(\ref{Rint_N3_Nf1}). In this formulation every configuration of link
variables $s(l)$ and $r(l)$ has non-zero Boltzmann weight, thus allowing us to
use again the simple Metropolis update algorithm instead of more complicated
worm-like algorithms usually adopted to probe dual model formulations.
The values of the function $Q_3(n,p)$ are computed beforehand and stored
in an array, which is then used in simulations.
An extra benefit of the absence of the triality condition is the possibility to
calculate a correlation function as the expectation value of a product of
one-site observables instead of a product over the path connecting the
sources (see Section~4).

Let us mention that our dual formulation allows to simulate all $SU(N)$ models 
on equal footing: the only difference between different $N$ is encoded in the
function $Q_N(n,p)$ which, as said above, can be computed prior to simulations.
Thus, the present approach can be easily extended to any values of $N$.
The investigation of the large-$N$ limit is interesting
  {\it per s\'e} since it can shed light on the large-$N$ QCD phase diagram
  at finite density, which could exhibit novel phases, such as quarkyonic
  matter~\cite{McLerran:2007qj,Hollowood:2012nr,Philipsen:2019qqm}. Moreover, it
  can reveal interesting connections between glueball states in $SU(N)$ gauge
  models for large $N$ and the constituents of dark matter~\cite{Huang:2020mso}.

For simulations above $h=0.01$, we performed $2 \cdot 10^5$ thermalization
updates, and then made measurements every 100 whole lattice updates, collecting
a statistic of $10^5$ measurements. To estimate statistical error, a jackknife
analysis was performed at different blocking over bins with size varying from
10 to $10^4$. 

When we move to smaller values of $h$, we see that transition probabilities
of the one-spin Metropolis update become smaller. To counter this, while
keeping the simulation time reasonable, we used the combination of a multihit
update with an update skipping procedure: we go through each link variable
and attempt $n_{\rm hit}$ Metropolis updates, but before each attempt we
calculate the {\it a priori} probability that the update is rejected and
generate the number $n_{\rm rej}$ of rejects before the first
acceptance (from the corresponding geometric distribution); in this way we
can skip $n_{\rm rej}$ updates and therefore
save on the number of updates left to perform on the current link.
Then, if we still need to do some updates, we choose the update for the
link using the probabilities calculated in presumption that the update is
accepted.
Using this technique allows us to vary $n_{\rm hit}$ for different simulation
parameters in the range 5 - 2000, while keeping the simulation time constant
for a fixed acceptance rate.

Below $h=0.01$ we performed $2 \cdot 10^5$ thermalization updates,
with measurements taken every 50 whole lattice updates,
collecting a statistics of $10^6$ measurements.
The bins size in the jackknife analysis varied from 10 to $10^5$.

\subsection{Critical behavior}

A clear indication of the presence of different phases can be seen from
the inspection of the distribution and the scatter plot of the Monte
Carlo equilibrium values for the absolute value of the magnetization near a
transition value of $\beta$: for some choice of the parameters $h$ and $\mu$,
we observe two separate peaks and two separate spots, respectively, which is
suggestive of a first order transition; for other choices we see instead
just one single peak and a single spot, respectively, as expected for a
second order transition or a crossover.

\begin{figure}[htb]
\centering
{
\hfill
\includegraphics[width=0.48\textwidth]{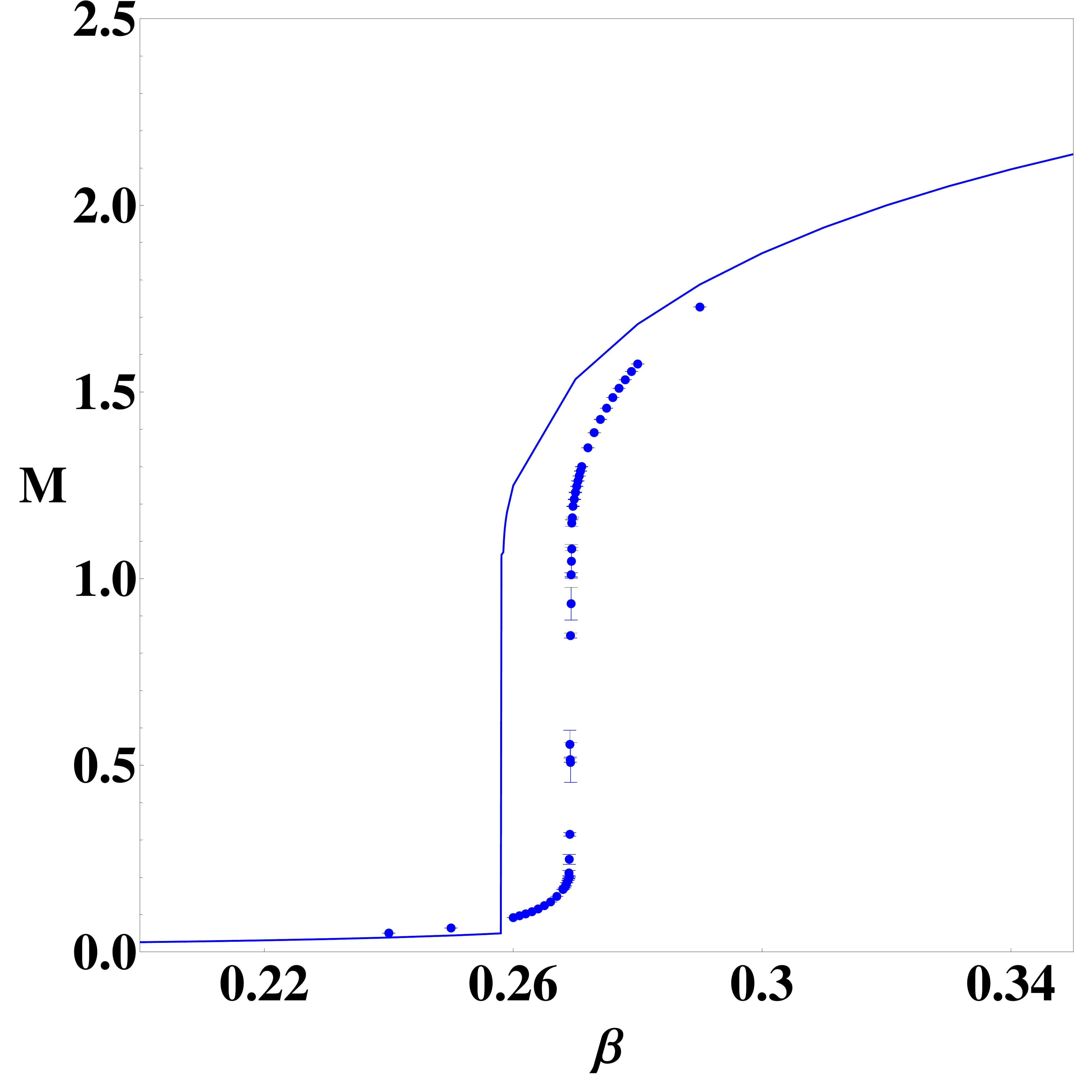}
\hfill
\includegraphics[width=0.48\textwidth]{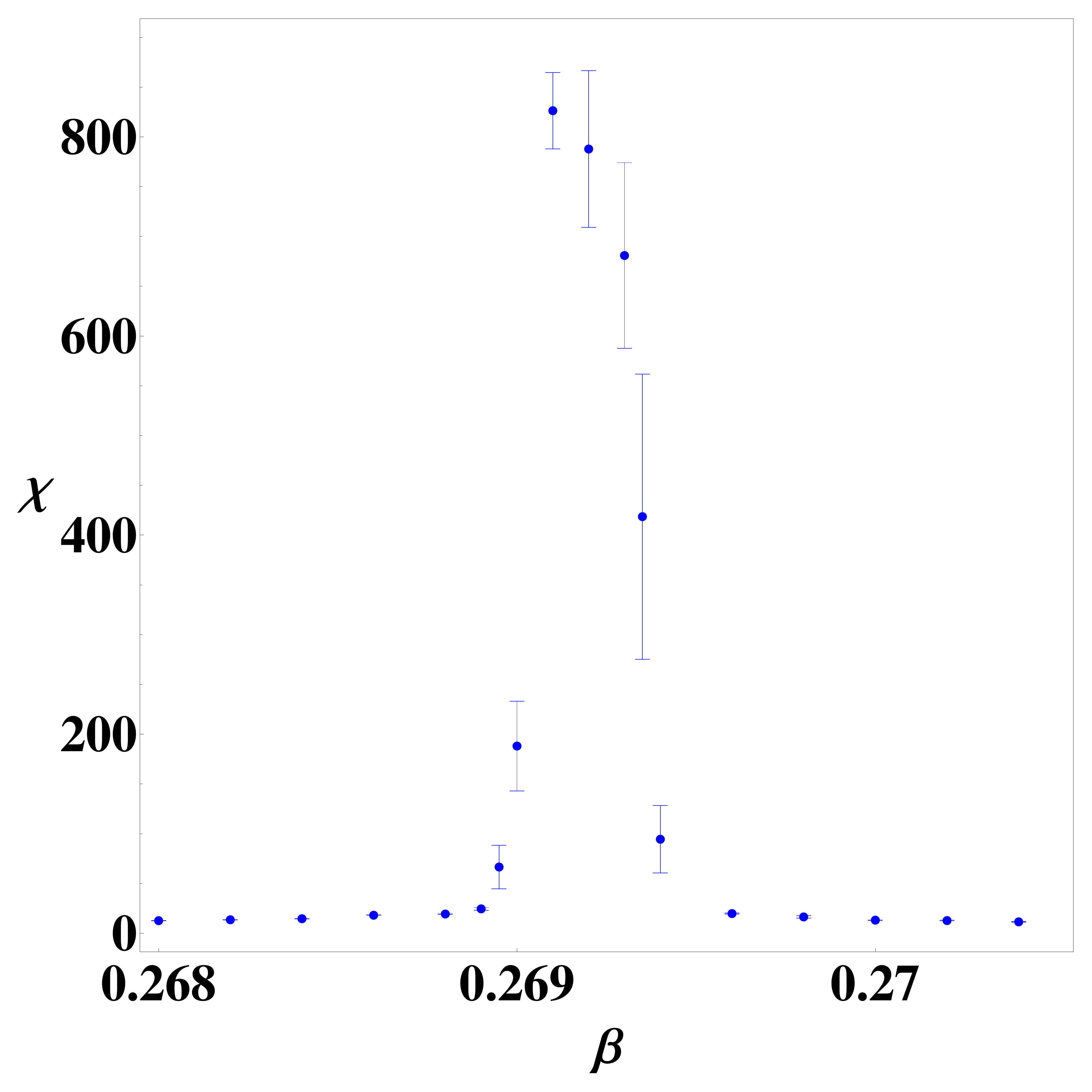}
\hfill
}
\caption{Magnetization (left) and magnetization susceptibility (right)
  {\it versus} $\beta$ at $h=0.01$ and $\mu=0$ on a $16^3$ lattice, in the
  vicinity of a first order phase transition. The solid lines represent
  the mean-field estimates.} \label{fig:M_Chi_I_Order}
\end{figure}

\begin{figure}[htb]
\centering
{
\hfill
\includegraphics[width=0.48\textwidth]{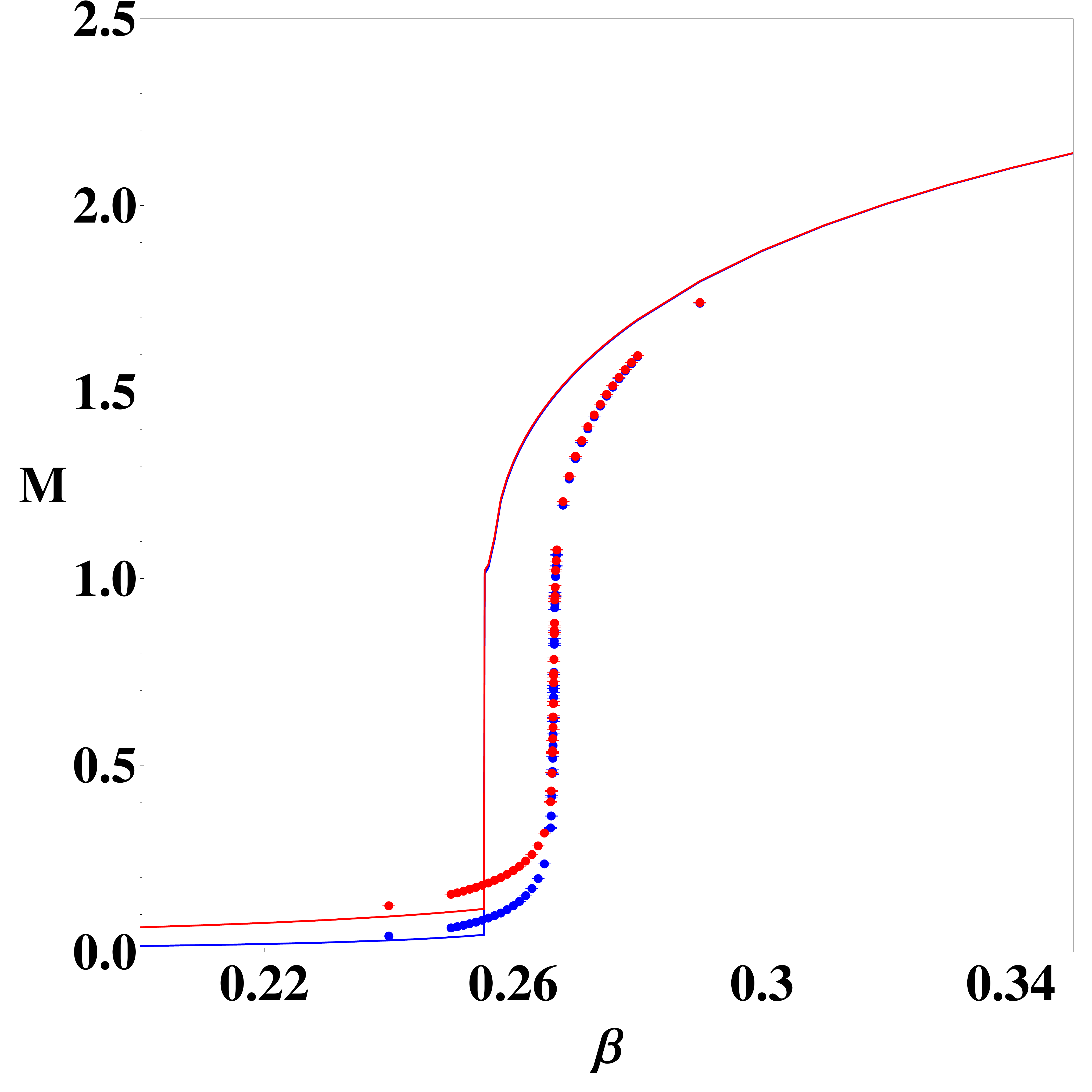}
\hfill
\includegraphics[width=0.48\textwidth]{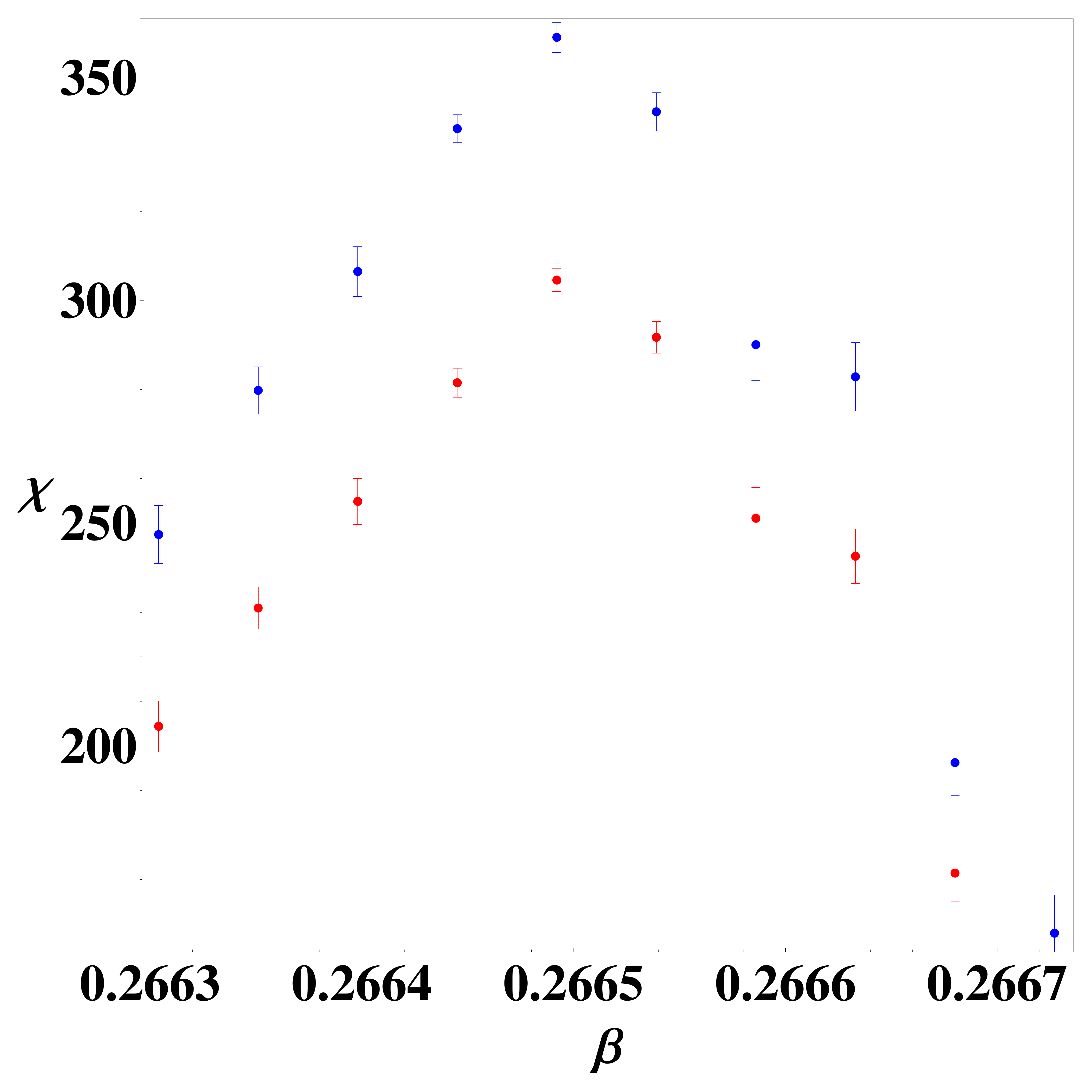}
\hfill
}
\caption{The same as Fig.~\ref{fig:M_Chi_I_Order} for $h=0.01$ and
  $\mu=0.9635$, where a second order phase transition occurs.
  Lines and points in blue correspond to the magnetization, while those in red
  correspond to the conjugate magnetization.}\label{fig:M_Chi_II_Order}
\end{figure}

\begin{figure}[htb]
\centering
{
\hfill
\includegraphics[width=0.48\textwidth]{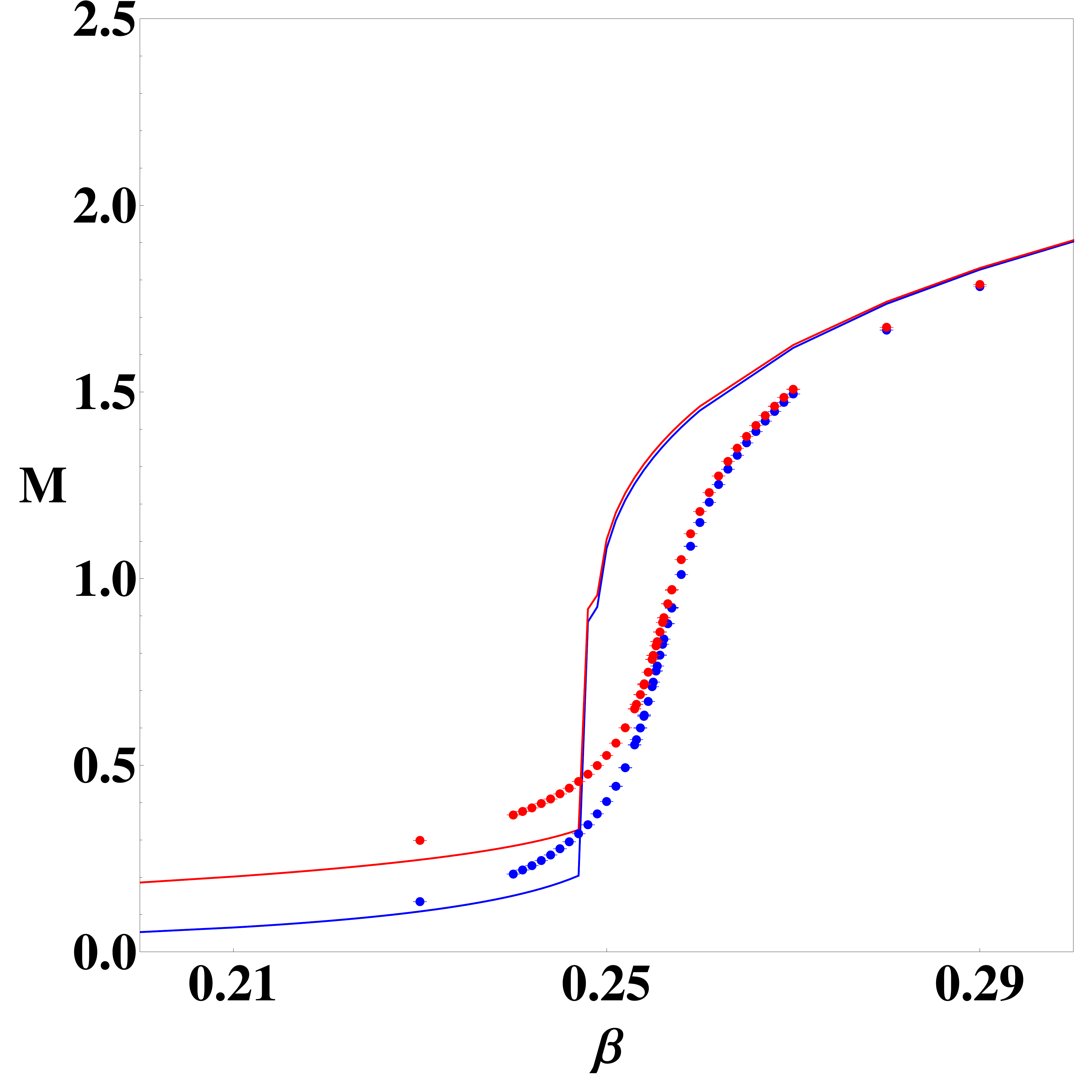}
\hfill
\includegraphics[width=0.48\textwidth]{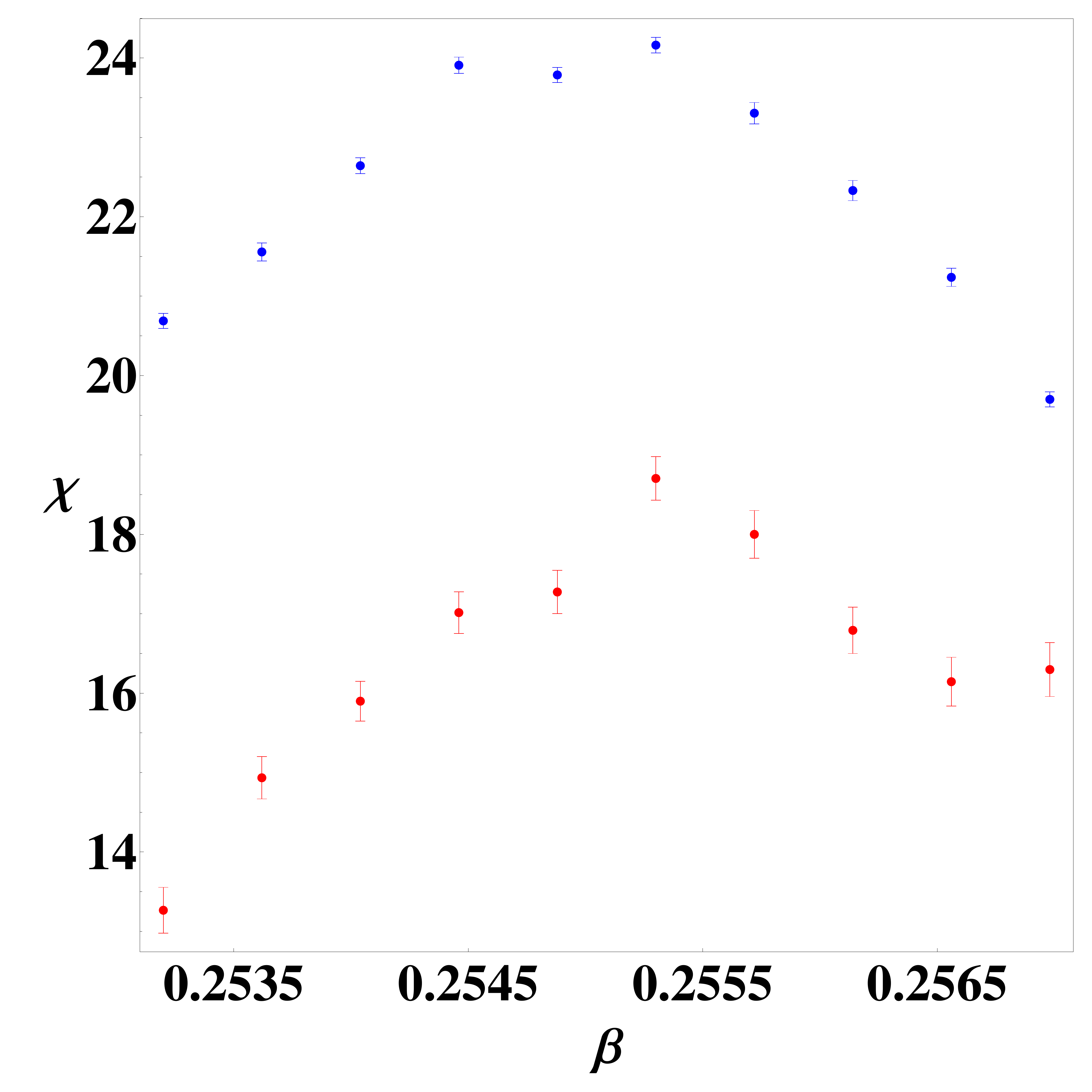}
\hfill
}
\caption{The same as Fig.~\ref{fig:M_Chi_II_Order} for $h=0.01$ and
  $\mu=2$, where a crossover transition occurs.}\label{fig:chi-M-Crossover}
\end{figure}

Another indication comes from the behavior of the magnetization and its
susceptibility {\it versus} $\beta$ near the transition for different
values of $(h, \mu)$. Figs.~\ref{fig:M_Chi_I_Order}-\ref{fig:chi-M-Crossover}
show that, when $h$ is kept fixed and $\mu$ is varied from $\mu=0$ to $\mu=2$,
the transition softens, the ``jump'' of the magnetization at the
pseudo-critical $\beta$ becoming less steep and the peak of the susceptibility
less pronounced, which suggests that different regimes are being explored.
According to the finite-size scaling analysis discussed below,
the three regimes represented in
Figs.~\ref{fig:M_Chi_I_Order}-\ref{fig:chi-M-Crossover} correspond
to first order, second order and crossover, respectively.
 
To characterize in a quantitative way the phase structure of the model in the
$(h, \mu)$-plane, we performed a standard finite-size scaling analysis on the
peak value of the magnetization susceptibility $\chi$.
Although the simulation algorithm does not prevent us from
  considering arbitrarily large lattice sizes, we limit ourselves to the results
  presented here, obtained by high-statistic simulations on lattices with linear
  sizes $L$ = 10 and 16. Larger lattices were also studied for some choices of
  the couplings to investigate the behavior of the correlation functions. 
  We do not quote those results here to keep a uniform treatment of the whole
  parameter space. These results will appear in a separate paper. The reason
  for using relatively small lattice sizes is twofold: on one side, we had to
  explore a wide region in a three-parameter space and simulating many volumes
  for each point in the parameter space would have been too expensive; on the
  other side, drawing a precision phase diagram for this model is beyond the
  scope of the present work, whose main aim is rather to show the effectiveness
  of the suggested model.
We determined $\chi$ for several $\beta$ values in the
transition region and fitted them to a Lorentzian, thus getting the position
of the peak, which gives the pseudocritical coupling $\beta_{\rm pc}$, and its
height. Comparing the dependence of the peak height on the lattice size $L$
with the scaling law
\begin{equation}
\label{sp_Height_scaling}
\chi_L(\beta_{\rm pc}) = A L^{\gamma/\nu}\;,
\end{equation}
we estimated the critical-index ratio $\gamma/\nu$ and collected all
our determinations, as many as 202, in 
Table~\ref{table:beta_critic_and_gamma_over_nu_versus_h_nu}.
More specifically, for each $(h,\mu)$ pair we extracted
  $\gamma/\nu$ by the following formula:
\begin{equation}
\dfrac{\gamma}{\nu} = \log_{\frac{16}{10}} \left( \dfrac{\chi_{16}
(\beta_{\rm pc})}{\chi_{10}(\beta_{\rm pc})}\right)  
\end{equation}
and assigned to each determination a statistical error calculated by
standard error propagation.

\begin{landscape}
\begin{table}
\begin{tiny}   
\begin{minipage}{0.4\textwidth}
         
\begin{tabular}{| c | c | c | c |}
\hline \hline
$h$  		& $\mu$			   & $\beta_{\rm pc}$   & $\gamma/\nu$ \\
\hline
0.001		&3.0&0.269041(7)	&2.83(6)\\
	 		&3.2&0.267843(5)	&2.50(11)\\
	 		&3.3&0.267156(7)	&2.14(5)\\
	 		&3.35&0.266782(5)	&2.06(4)\\
	 		&3.36&0.266708(1)	&1.96(2)\\
	 		&3.365&0.266669(3)	&1.93(2)\\
	 		&3.375&0.266596(4)	&1.88(3)\\
	 		&3.4&0.266390(4)	&1.79(4)\\
	 		&3.5&0.265563(5)	&1.39(4)\\
			&4.0&0.26016(4)		&0.07(9)\\
	 		&5.0&0.2413(2)			&0.01(3)\\

\hline
0.002		&2.3&0.26904(2)&2.72(10)\\
	 		&2.4&0.268487(4)	&2.48(11)\\
	 		&2.5&0.267861(5)	&2.54(3)\\
	 		&2.55&0.267526(4)	&2.35(3)\\
	 		&2.6&0.267172(6)	&2.24(6)\\
	 		&2.625&0.266992(5)	&2.15(3)\\
	 		&2.65&0.266809(5)	&2.03(4)\\
	 		&2.8&0.26558(2)		&1.49(10)\\
	 		&3.0&0.26370(3)		&0.61(13)\\
	 		&4.0&0.2485(3)		&-0.07(6)\\
	 
\hline
0.003		&2&0.26837(2)		&2.56(28)\\
	 		&2.2&0.267102(6)	&2.24(4)\\
	 		&2.25&0.266724(3)	&1.97(2)\\
	 		&2.275&0.266537(5)	&1.92(3)\\
	 		&2.3&0.266339(3)	&1.77(2)\\
	 		&2.5&0.264611(7)	&0.99(3)\\

\hline
0.004		&1.6&0.268926(9)	&2.64(18)\\
	 		&1.8&0.267786(4)	&2.48(3)\\
	 		&1.88&0.267257(4)	&2.22(3)\\
	 		&1.9&0.267116(6)	&2.18(4)\\
	 		&1.95&0.266758(3)	&2.02(3)\\
	 		&1.96&0.266685(3)	&1.98(3)\\
			&1.97&0.266608(4)	&1.94(2)\\
	 		&1.975&0.266574(5)	&1.87(3)\\
	 		&2&0.26637(2)		&1.63(22)\\
	 		&3&0.2542(1)		&0.08(6)\\ 
\hline
0.005		&1.5&0.268147(4)	&2.58(4)\\
	 		&1.7&0.266877(4)	&2.07(3)\\
	 		&1.8&0.266137(4)	&1.69(3)\\
	 		&1.85&0.265729(2)	&1.50(1)\\
	 		&2&0.264402(5)		&0.98(4)\\
	 		&2.5&0.25850(5)		&0.12(7)\\
\hline
0.007		&0.8&0.26947(4)		&2.50(46)\\
	 		&1&0.268710(7)		&2.46(14)\\
	 		&1.2&0.267686(3)	&2.52(5)\\
	 		&1.3&0.267085(4)	&2.23(3)\\
	 		&1.4&0.266403(4)	&1.92(5)\\
	 		&1.5&0.265659(4)	&1.53(3)\\
	 		&2&0.26066(4)		&0.21(7)\\
	 		&3&0.2429(2)		&0.16(4)\\
\hline	 \hline
\end{tabular}
\end{minipage}
\begin{minipage}[c]{0.42\textwidth}
\begin{tabular}{| c | c | c | c |}
\hline \hline
$h$  		& $\mu$			   & $\beta_{\rm pc}$   & $\gamma/\nu$ \\
\hline	 
0.008		&0&0.2705(2)		&3.02(13)\\
		 	&0.8&0.268774(9)	&2.84(5)\\
		 	&1&0.26787(1)		&2.60(11)\\
		 	&1.2&0.266733(1)	&2.06(1)\\
		 	&1.3&0.266044(2)	&1.69(1)\\
	 		&1.35&0.265672(3)	&1.50(2)\\
		 	&1.5&0.264432(8)	&0.99(5)\\
			&2.5&0.25(9)		&0.04(4)\\
\hline
0.01		&0		&0.26916(2)	&3.19(60)\\
			&0.01	&0.26923(1)	&2.87(6)\\
			&0.2	&0.26913(2)	&2.92(8)\\
			&0.4	&0.26872(1)	&2.91(10)\\
			&0.7	&0.26780(2)	&2.50(10)\\
			&0.8	&0.26734(2)	&2.21(16)\\
			&0.9	&0.266844(1)&2.14(6)\\
			&0.95	&0.26656(1)	&2.04(4)\\
			&0.96	&0.266504(5)&2.02(4)\\
			&0.963	&0.26648(1)	&1.95(3)\\
			&0.96325&0.26649(1)	&1.95(3)\\
			&0.9635	&0.26649(1)	&1.94(4)\\
			&0.964	&0.266473(4)&1.95(2)\\
			&0.96425&0.26648(1)	&1.97(3)\\
			&0.9645	&0.26647(1)	&1.98(3)\\
			&0.965	&0.26646(1)	&1.92(3)\\
			&0.967	&0.26646(1)	&1.92(3)\\
			&0.97	&0.26644(1)	&1.92(3)\\
			&0.98	&0.26638(1)	&1.86(5)\\
			&1		&0.26626(1)	&1.82(4)\\
			&1.		1&0.26558(1)&1.44(4)\\
			&1.5	&0.26204(1)	&0.40(5)\\
			&2		&0.25543(9)	&0.10(5)\\
\hline
0.0125		&0		&0.26791(1)	&2.63(7)\\
			&0.0125	&0.26789(1)	&2.55(6)\\
			&0.125	&0.26785(1)	&2.56(4)\\
			&0.2	&0.26776(1)	&2.57(5)\\
			&0.3	&0.26758(1)	&2.52(4)\\
			&0.4	&0.26735(1)	&2.43(4)\\
			&0.5	&0.267032(4)&2.32(3)\\
			&0.6	&0.266639(4)&2.06(3)\\
			&0.65	&0.26641(1)	&1.93(4)\\
			&0.7	&0.26617(1)	&1.83(4)\\
			&0.8	&0.26562(1)	&1.54(3)\\
			&0.9	&0.26499(1)	&1.27(7)\\
			&0.95	&0.26462(1)	&1.08(4)\\
			&0.97	&0.26446(1)	&0.96(4)\\
			&1		&0.26425(1)	&0.99(6)\\
			&1.25	&0.26205(2)	&0.44(8)\\
			&		&			&		\\
			&		&			&		\\
			&		&			&		\\
			&		&			&		\\
\hline \hline
\end{tabular}
\end{minipage}
\begin{minipage}[c]{0.42\textwidth}
\begin{tabular}{| c | c | c | c |}
\hline \hline
$h$  		& $\mu$			   & $\beta_{\rm pc}$   & $\gamma/\nu$ \\
\hline
0.014		&0		&0.26712(1)	&2.39(8)\\
			&0.014	&0.267132(4)&2.37(3)\\
			&0.14	&0.26706(1)	&2.32(3)\\
			&0.16	&0.26702(1)	&2.32(4)\\
			&0.2	&0.26697(1)	&2.33(3)\\
			&0.3	&0.26678(1)	&2.24(4)\\
			&0.4	&0.266513(5)&2.11(4)\\
			&0.5	&0.266164(5)&1.83(3)\\
			&0.6	&0.265716(5)&1.63(2)\\
			&0.8	&0.26458(1)	&1.04(4)\\
			&0.9	&0.26387(1)	&0.86(3)\\
			&1		&0.26307(1)	&0.61(3)\\
			&1.2	&0.26117(1)	&0.31(4)\\
			&1.4	&0.25884(5)	&0.12(4)\\
			&2		&0.2491(2)	&-0.04(4)\\
\hline
0.015		&0		&0.26660(1)&2.13(6)\\
			&0.015	&0.26660(1)&2.19(3)\\
			&0.15	&0.26651(1)&2.04(5)\\
			&0.2	&0.26644(1)&2.06(4)\\
			&0.3	&0.26625(1)&1.94(3)\\
			&0.35	&0.26613(1)&1.91(3)\\
			&0.4	&0.26596(1)&1.82(3)\\
			&0.45	&0.265778(4)&1.71(2)\\
			&0.5	&0.26556(1)&1.44(2)\\
			&0.6	&0.26510(1)&1.34(3)\\
\hline
0.0155		&0		&0.26635(1)&1.99(3)\\
			&0.0155	&0.26635(1)&2.05(8)\\
			&0.05	&0.26635(1)&2.01(3)\\
			&0.07	&0.266332(4)&2.02(3)\\
			&0.08	&0.266312(5)&1.95(2)\\
			&0.085	&0.26632(1)&1.93(4)\\
			&0.09	&0.26632(1)&2.04(3)\\
			&0.1	&0.26630(1)&1.95(3)\\
			&0.155	&0.26625(1)&1.93(4)\\
			&0.3	&0.26597(1)&1.86(4)\\
			&0.7	&0.26423(1)&1.02(3)\\
			&1		&0.26193(2)&0.44(2)\\
\hline
0.0156		&0		&0.266303(4)&1.94(3)\\
			&0.0156	&0.266292(3)&2.01(2)\\
			&0.02	&0.266295(4)&1.96(2)\\
			&0.025	&0.26629(1)&1.98(4)\\
			&0.0275	&0.26631(1)&1.95(2)\\
			&0.03	&0.266293(5)&1.94(3)\\
			&0.156	&0.26620(1)&1.92(3)\\
			&0.2	&0.266121(1)&1.90(4)\\
			&0.4	&0.265623(4)&1.61(3)\\
			&0.6	&0.26475(1)&1.22(3)\\
			&0.8	&0.263497(4)&0.74(3)\\
			&		&			&		\\
			&		&			&		\\
			&		&			&		\\
\hline \hline
\end{tabular}
\end{minipage}
\begin{minipage}[c]{0.42\textwidth}
\begin{tabular}{| c | c | c | c |}
\hline \hline
$h$  		& $\mu$			   & $\beta_{\rm pc}$   & $\gamma/\nu$ \\
\hline
0.0157		&0		&0.266252(4)&2.01(3)\\
			&0.0157	&0.26625(1)	&1.97(3)\\
			&0.02	&0.266247(1)&1.97(1)\\
			&0.03	&0.26624(1)	&1.99(3)\\
			&0.05	&0.266235(1)&1.98(3)\\
			&0.07	&0.26623(1)	&1.92(3)\\
			&0.1	&0.26621(1)	&1.91(3)\\
			&0.2	&0.266072(4)&1.80(2)\\

\hline
0.0158		&0		&0.266198(3)&1.95(2)\\
			&0.01	&0.26620(1)	&1.89(3)\\
			&0.0158	&0.26619(1)	&1.94(4)\\
			&0.02	&0.266197(2)&1.91(2)\\
			&0.04	&0.266191(4)&1.95(3)\\
			&0.06	&0.266180(4)&1.93(2)\\
			&0.08	&0.26617(1)	&1.90(2)\\
			&0.1	&0.266151(4)&1.88(2)\\
			&0.158	&0.266085(4)&1.84(2)\\

\hline 
0.01585		&0		&0.26615(1)	&1.94(4)\\
			&0.01585&0.266171(4)&1.96(3)\\
			&0.02	&0.26617(1)	&1.96(4)\\
			&0.025	&0.26617(1)	&1.88(4)\\
			&0.03	&0.26616(1)	&1.95(3)\\
			&0.06	&0.26615(1)	&1.87(4)\\
			&0.1	&0.26613(1)	&1.92(3)\\
			&0.2	&0.26600(1)	&1.81(3)\\
\hline
0.0159		&0		&0.266157(4)&1.90(2)\\
			&0.0159	&0.26615(1)	&1.93(4)\\
			&0.02	&0.266150(5)&1.88(2)\\
			&0.04	&0.266152(4)&1.91(2)\\
			&0.06	&0.26613(1)	&1.93(4)\\
			&0.08	&0.266120(4)&1.91(2)\\
			&0.1	&0.26611(1)	&1.85(3)\\
			&0.159	&0.266039(3)&1.87(2)\\
			&0.2	&0.26598(1)	&1.79(3)\\
\hline
0.016		&0		&0.26616(1)	&1.90(4)\\
			&0.016	&0.26609(1)	&1.84(4)\\
			&0.16	&0.265976(3)&1.87(2)\\
\hline
0.017		&0		&0.26557(1)	&1.69(4)\\
			&0.017	&0.26558(1)	&1.60(4)\\
			&0.17	&0.26545(1)	&1.56(3)\\
\hline
0.018		&0.018	&0.26508(1)	&1.42(4)\\
\hline
0.02		&0		&0.26404(9)	&0.87(1)\\
			&0.01	&0.26404(1)	&1.00(4)\\
			&0.1	&0.26398(1)	&0.93(4)\\
			&1		&0.25865(6)	&0.14(3)\\
\hline
0.025		&0		&0.26156(1)	&0.40(2)\\
0.03		&0		&0.25913(3)	&0.10(2)\\
0.04		&0		&0.25466(4)	&0.01(2)\\
			&		&			&		\\
			&		&			&		\\
			&		&			&		\\
\hline \hline         
\end{tabular}
\end{minipage}

\caption{Summary of the determinations of $\beta_{\rm pc}$ and $\gamma/\nu$
  for different values of $h$ and $\mu$.}
\label{table:beta_critic_and_gamma_over_nu_versus_h_nu}
\end{tiny}
\end{table}
\end{landscape}

We can see that, within uncertainties, the values of $\gamma/\nu$ are
spread in a range between 3., which implies a first order transition, and
zero, which holds for crossover, passing through the second order
3-dimensional Ising value, $\gamma/\nu=1.9638(8)$~\cite{PELISSETTO2002549}.
These sparse values of $\gamma/\nu$ are evidently an artifact of the relatively
small lattice sizes we could simulate. If we could approach the thermodynamic
limit, we would see that values of $\gamma/\nu$ concentrate around the values
of 3. (first order), 1.9638(8) (second order in the 3-dimensional Ising class)
and 0. (crossover). This is expected since we know that at $\mu=0$ in the
pure gauge limit of QCD or for heavy enough quark masses there is a whole
region of first order deconfinement transitions in the $m_{\rm u,d}$-$m_s$ plane
(the famous Columbia plot), delimited by a line of second order critical
points in the 3-dimensional Ising class~\cite{Karsch:2000xv}: thereafter, for
lower quark masses, the crossover region is met. In the simulations of our
effective Polyakov loop model at non-zero density toward the thermodynamic
limit we should see the continuation of the line of second order critical
points to non-zero values of the chemical potential. For the lattice volumes
considered in our study, we are not able to make a clear-cut
assignment of each choice of the parameters $h$ and $\mu$ to one of the three
transition regions. Using the determination of $\gamma/\nu$, we tried anyhow to
make this assignment, extending and modifying the three possible options
(first order, second order and crossover) as in
Table~\ref{table:gamma_over_nucolor}. This makes no sense in the
thermodynamic limit, but can be helpful in the present context.
In Table~\ref{table:gamma_over_nucolor} we introduced a color code, to
help identifying at a glance in which of these regions a given parameter pair
$(h,\mu)$ falls.

\begin{table}[htb]
\centering 
\begin{tabular}{ | c | c | c |}
    \hline \hline
$\gamma/ \nu$  					& color			   &phase\\
\hline \hline
$\gamma/\nu\geq $3 				&green			   &first order\\
2.50 $\leq\gamma/\nu<$ 3		&light green	   &more first order than second order\\
1.98 $<\gamma/\nu<$ 2.50		&yellow			   &more second order than first order\\
1.94 $\leq\gamma/\nu\leq$ 1.98	&red			   &very close to second order\\
1.85 $\leq\gamma/\nu<$ 1.94 	&brown			   &more second order than crossover\\
0.3  $\leq\gamma/\nu<$ 1.85		&magenta		   &more crossover than second order\\
0    $\leq\gamma/ \nu<$ 0.3		&blue			   &crossover\\
\hline \hline
\end{tabular}
\caption{Color code used to characterize the different phases depending on 
  the value of $\gamma/\nu$. }
\label{table:gamma_over_nucolor}
\end{table}

\begin{figure}[htb]
\centering
{\includegraphics[width=1.0\textwidth]{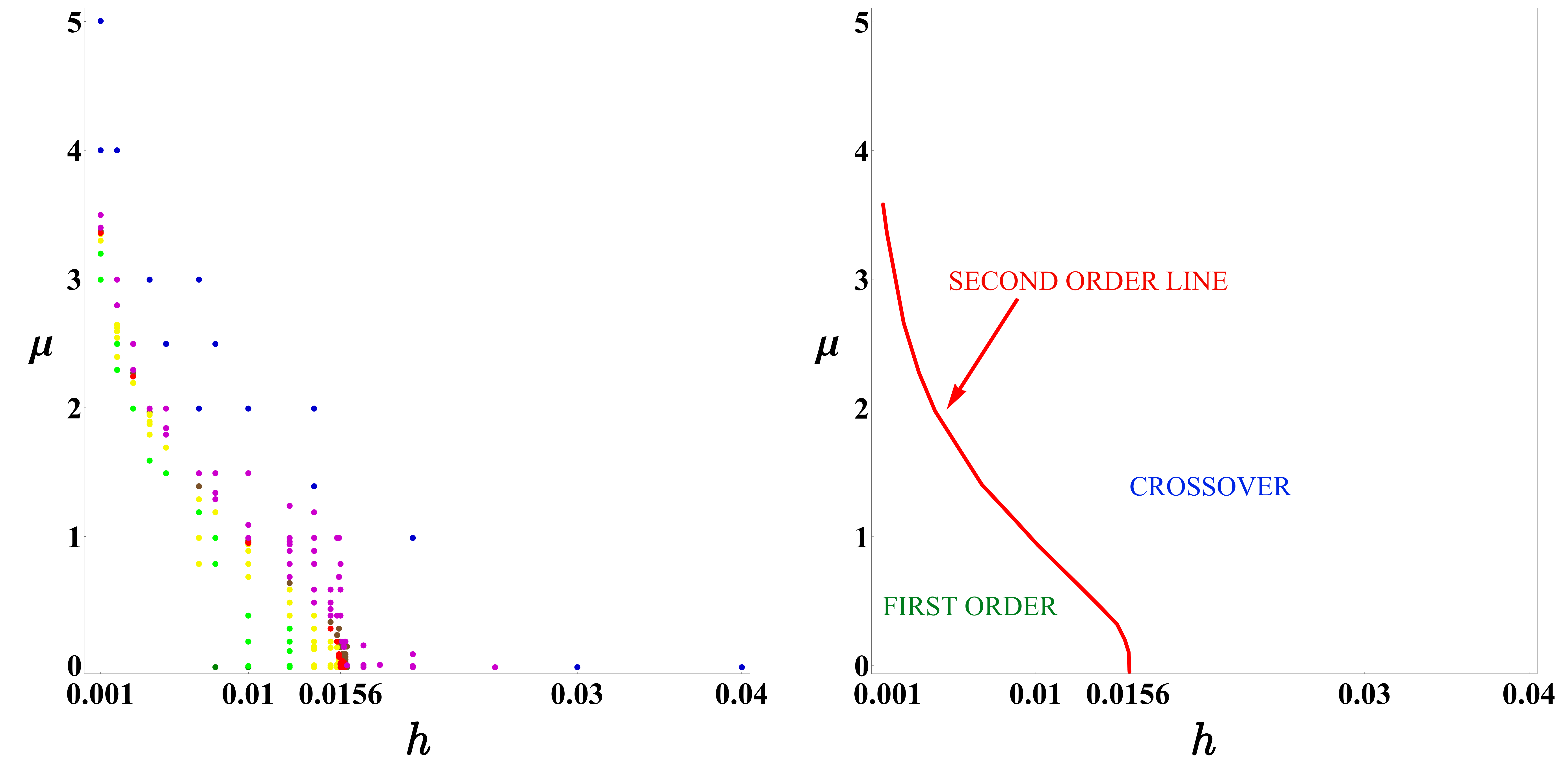}}
\caption{(Left) Assignment of each parameter pair $(h,\mu)$ to a transition
  region according to the color code of Table~\ref{table:gamma_over_nucolor}.
  (Right) Estimated phase diagram.}
\label{fig:Phase_diagram}
\end{figure}

\begin{figure}[htb]
\centering
{\includegraphics[width=0.64\textwidth]{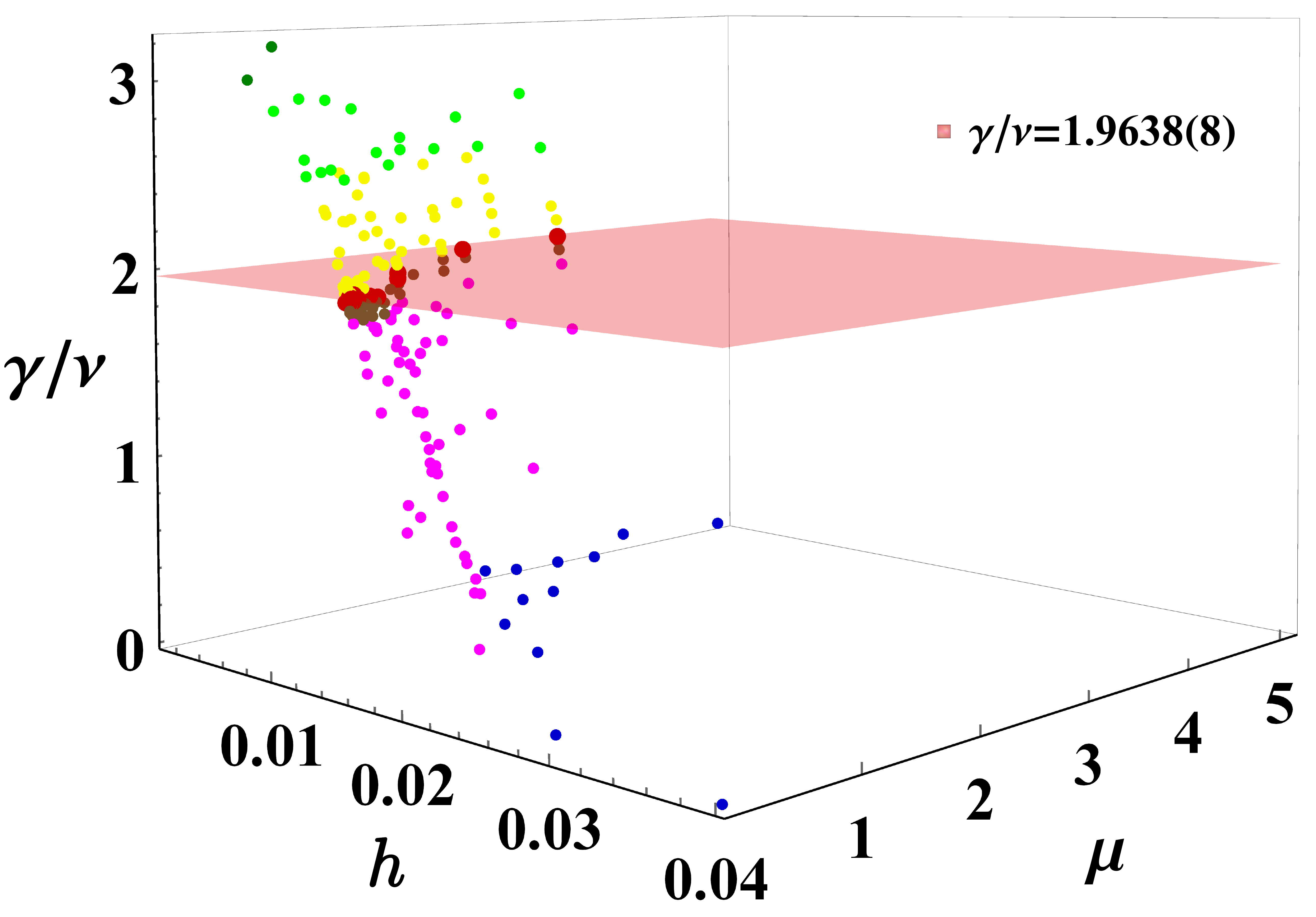}}
\caption{Values of the critical index ratio $\gamma/\nu$ for
  each considered choice of the pair $(h,\mu)$.
  The color code is as in Table~\ref{table:gamma_over_nucolor}. 
  The plane at $\gamma/\nu=1.9638(8)$ corresponds to the $\gamma/\nu$ value
  for a second order phase transition in the 3-dimensional Ising
  class~\cite{PELISSETTO2002549} .}\label{fig:gamma_over_nu_vs_h_mu}
\end{figure}

In Fig.~\ref{fig:Phase_diagram}(left) each parameter pair $(h,\mu)$ considered
in our simulations is represented by a colored dot in the $(h,\mu)$ plane,
according to the color code defined in Table~\ref{table:gamma_over_nucolor},
allowing us to sketch a tentative phase diagram in the right panel
of the same figure. A different visualization of the general phase diagram
is presented in Fig.~\ref{fig:gamma_over_nu_vs_h_mu}, where in a $3d$ plot
$\gamma/\nu$ values are reported in correspondence of each $(h,\mu)$ pair.

\begin{figure}[htb]
\centering
{\includegraphics[width=0.64\textwidth]{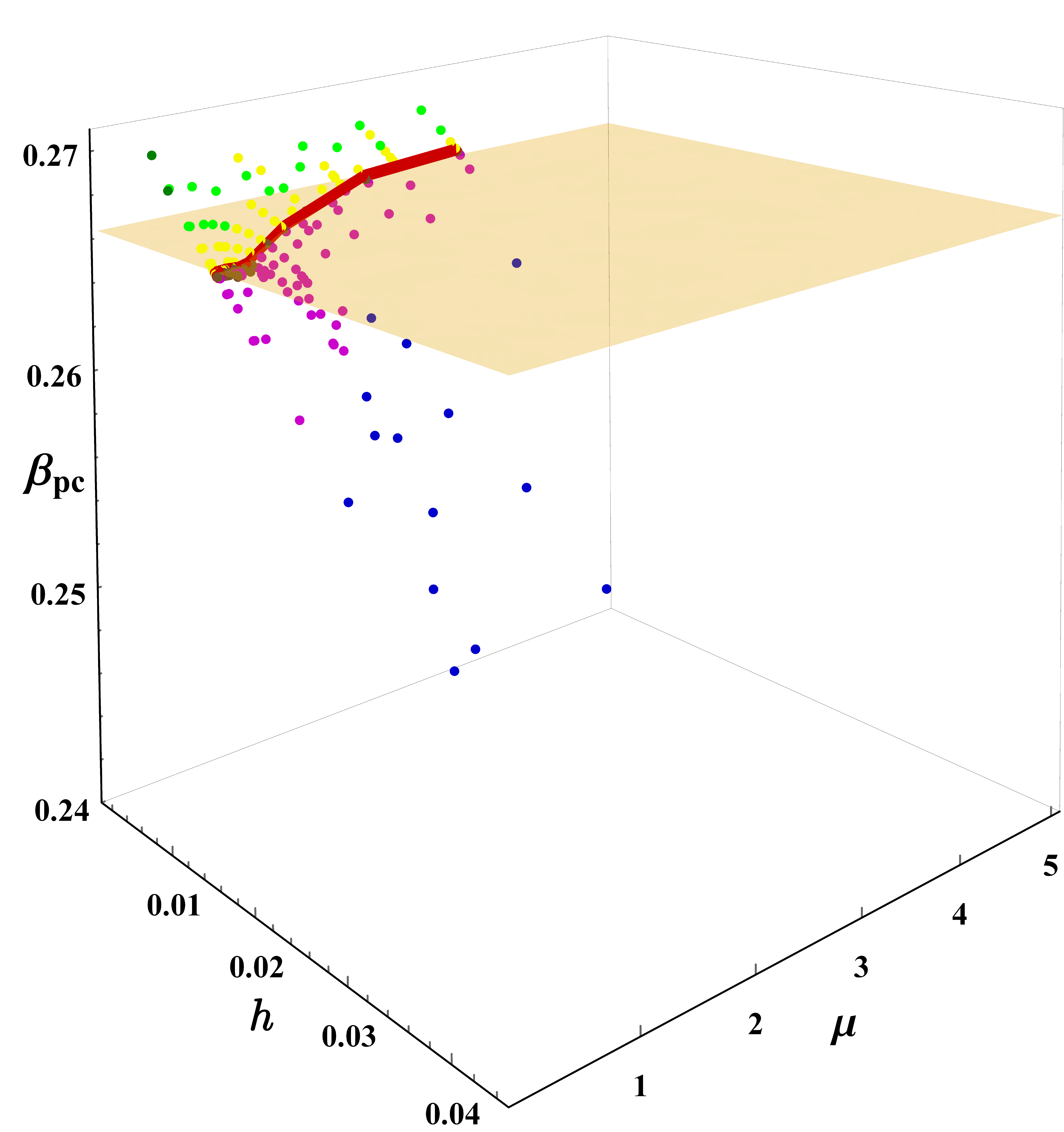}}
\caption{Values of $\beta_{\rm pc}$ for each considered choice of the pair
  $(h,\mu)$. The color code is as in Table~\ref{table:gamma_over_nucolor}. 
  For the sake of readability,
    red points have been connected by a broken line and the plane
    $\beta_{\rm pc}=0.266334$ was drawn.}\label{fig:beta_critical_vs_h_mu}
\end{figure}

Fig.~\ref{fig:beta_critical_vs_h_mu} shows the values of $\beta_{\rm pc}$
obtained for each considered pair $(h,\mu)$.
It is interesting to note that the variation of $\beta_{pc}$
  along the red line (close to the second order phase transition) is much
  smaller than the variation for all other points. Hence, red points lie
  approximately in one plane $\beta_{\rm pc}={\rm const}$, while all points
  associated with other phases lie on a curved surface.
This is also seen from Table~\ref{table:beta_critic_and_gamma_over_nu_very_close_to_second_order}, where we collected the values of $\beta_{\rm pc}$
close to the second order phase transition. 

\begin{table}[htb]
\centering 
\begin{tabular}{ | c | c | c | c |}
    \hline \hline
$h$  		& $\mu$			   & $\beta_{\rm pc}$   & $\gamma/\nu$ \\
\hline \hline
0.001&3.36&0.266708(1)&1.96(2)\\
0.003&2.25&0.266724(3)&1.97(2)\\
0.01&0.963&0.26648(1)&1.95(3)\\
0.01&0.96325&0.26649(1)&1.95(3)\\
0.01&0.9635&0.26649(1)&1.94(3)\\
0.01&0.964&0.266473(4)&1.95(2)\\
0.01&0.96425&0.26648(1)&1.97(3)\\
0.01&0.9645&0.26647(1)&1.98(3)\\
0.015&0.3&0.26625(1)&1.94(3)\\
0.0154&0.2&0.26624(1)&1.96(4)\\
0.0155&0.08&0.266312(5)&1.95(2)\\
0.0155&0.1&0.26630(1)&1.95(3)\\
0.0156&0.0&0.266303(4)&1.94(3)\\
0.0156&0.02&0.266295(4)&1.96(2)\\
0.0156&0.025&0.26629(1)&1.98(4)\\
0.0156&0.0275&0.26631(1)&1.95(2)\\
0.0156&0.03&0.266293(5)&1.94(3)\\
0.0157&0.0&0.26625(1)&1.97(3)\\
0.0157&0.02&0.266247(1)&1.97(1)\\
0.0157&0.05&0.266235(1)&1.98(3)\\
0.0158&0.0&0.266198(3)&1.95(2)\\
0.0158&0.04&0.266191(4)&1.95(3)\\
0.01585&0.0&0.26615(1)	&1.94(4)\\
0.01585&0.01585&0.266171(4)&1.96(3)\\
0.01585&0.02&0.26617(1)&1.96(3)\\
0.01585&0.03&0.26616(1)&1.95(3)\\
\hline \hline
\end{tabular}
\caption{Values of $\beta_{\rm pc}$ and $\gamma/\nu$ for $h$ and $\mu$ belonging
  to the region ``very close to second order''. 
  These values correspond to the bigger red points in
  Fig.~\ref{fig:beta_critical_vs_h_mu}.}
\label{table:beta_critic_and_gamma_over_nu_very_close_to_second_order}
\end{table}

We have not performed simulations in the absence of external field. To find the
critical value $\beta_{\rm g}$ at $h=0$, {\it i.e.} for the pure gauge
theory, and at $\mu=0.$, we performed a simple fit of the form
$\beta_{\rm c}(h) = \beta_{\rm g}
+ a h$, as suggested in~\cite{Philipsen12}. We obtained following values
$\beta_g=0.274991$, $a=-0.5568$. The value of $\beta_{\rm g}$ agrees very well
with the value quoted in the literature, $\beta_{\rm g}=0.274$ and reasonably
well with the mean-field result $\beta_{\rm g}=0.2615$. 

Another important question concerns the shape of the critical line shown in
the right panel of Fig.~\ref{fig:Phase_diagram} in the heavy-dense limit,
$h\to 0$, $\mu\to\infty$. From the data we have, one cannot make unambiguous
conclusions about its behavior. Nevertheless, data are well fitted by the
function $\mu_{\rm c}= - a \ln h +c - b h^2$, with $a=0.932$, $c=-3.1$, $b=2367$.
This shows that the line of second order phase transition might persist 
in the heavy-dense limit of QCD.

\subsection{Quark condensate and baryon density}

In this subsection we study the behavior of the quark condensate and the baryon
density in different phases and compare numerical results with mean-field
predictions. In the static approximation for the quark determinant one cannot
observe the phenomenon of spontaneous breaking of the chiral symmetry. Indeed,
even in the strong coupling region, using Eq.~(\ref{strong-coupling}), one can
easily obtain for the quark condensate $Q=0$ for all $\mu$ in the massless
limit $h=1$. The same result in this limit can be obtained within mean-field
approach and from numerical simulations which we performed for various values
of $\beta$ and $\mu$. Nevertheless, we think it might be instructive to see the
behavior of the condensate in the three regimes corresponding to first and
second order transitions and to crossover.
A more traditional observable to study in the effective Polyakov loop models is
the baryon density. In the dual formulation the baryon density $B$ and the
quark condensate $Q$ are given by Eqs.~(\ref{baryonden_dual})
and~(\ref{q_cond_dual}), respectively. We have computed the right-hand sides of
these equations both numerically and using the mean-field approach, for the same
values of $h$ and $\mu$. 

\begin{figure}[htb]
\centering
{
\hfill
\includegraphics[width=0.48\textwidth]{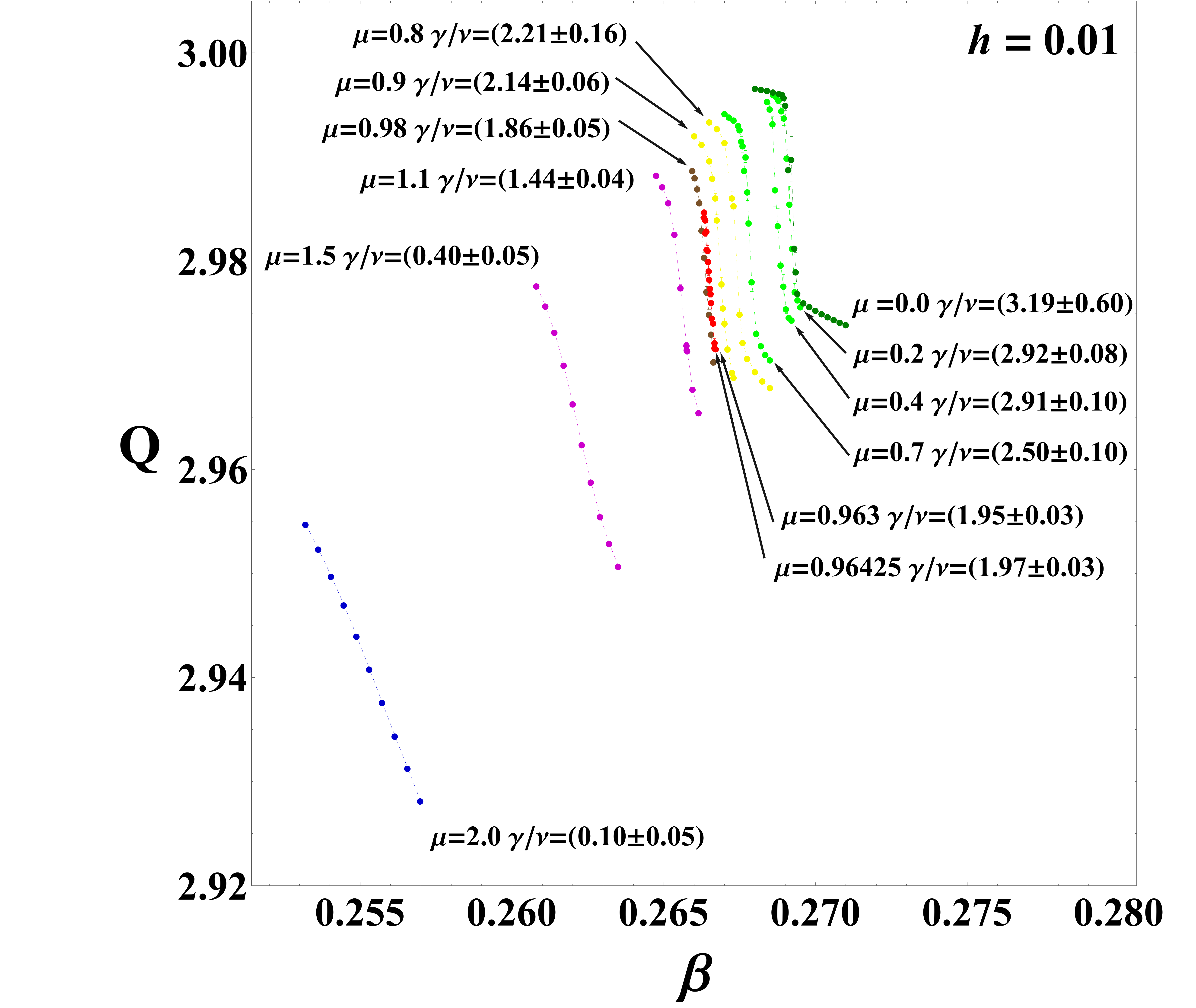}
\hfill
\includegraphics[width=0.48\textwidth]{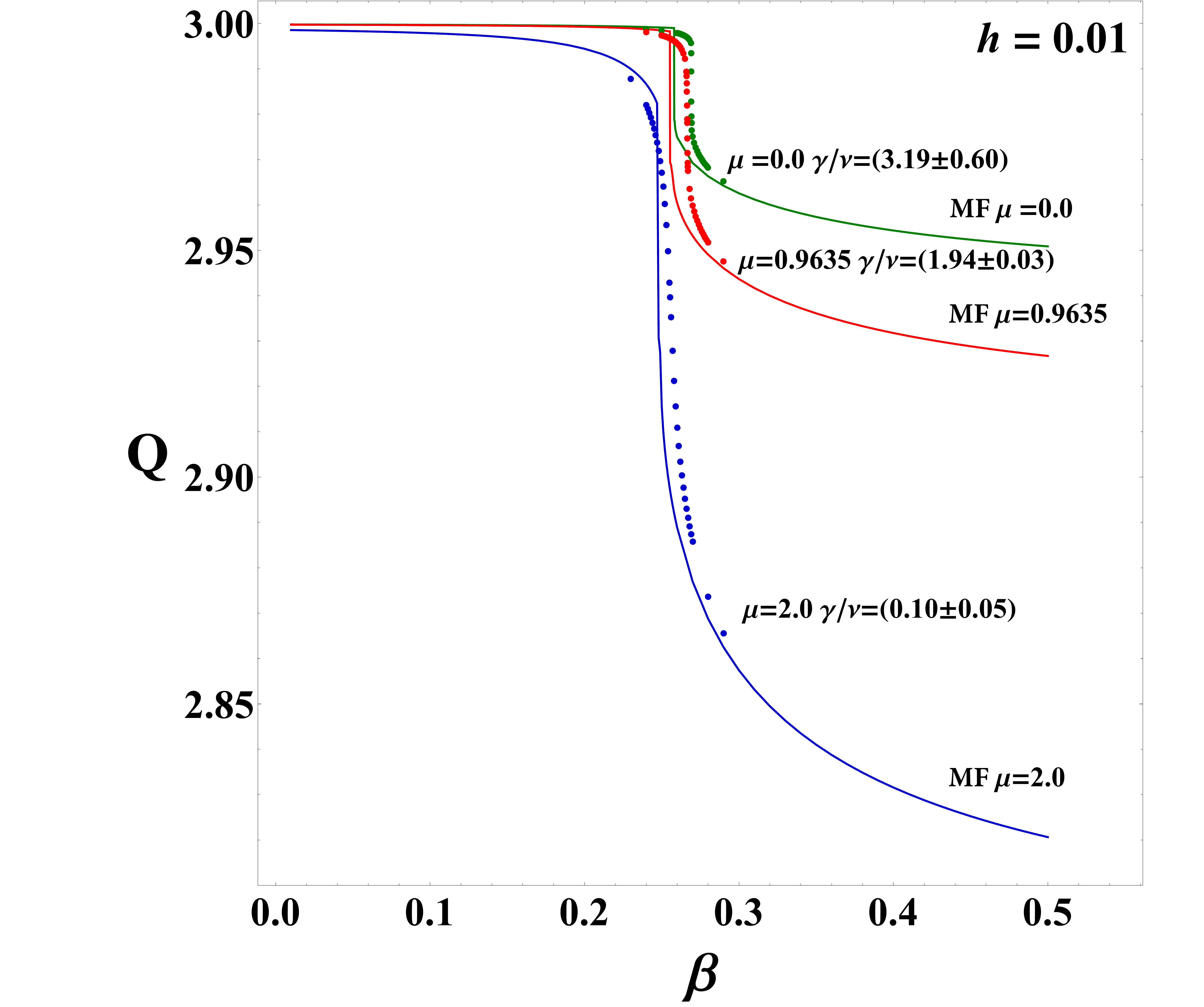}
\hfill
}
\caption{Behavior of the quark condensate versus $\beta$ for fixed $h=0.01$ and
  for different values of $\mu$ on a 16$^3$ lattice.
  Left panel: results of simulations in the
  vicinity of a phase transition. Right panel: comparison of mean-field
  analysis and numerical results in three different phase regimes. The color
  code of Table~\ref{table:gamma_over_nucolor} is used here and below to
  differentiate the phases of the system.}
\label{fig:2D_Quark_condensate_h0.01}
\end{figure}

\begin{figure}[htb]
\centering
{
\hfill
\includegraphics[width=0.48\textwidth]{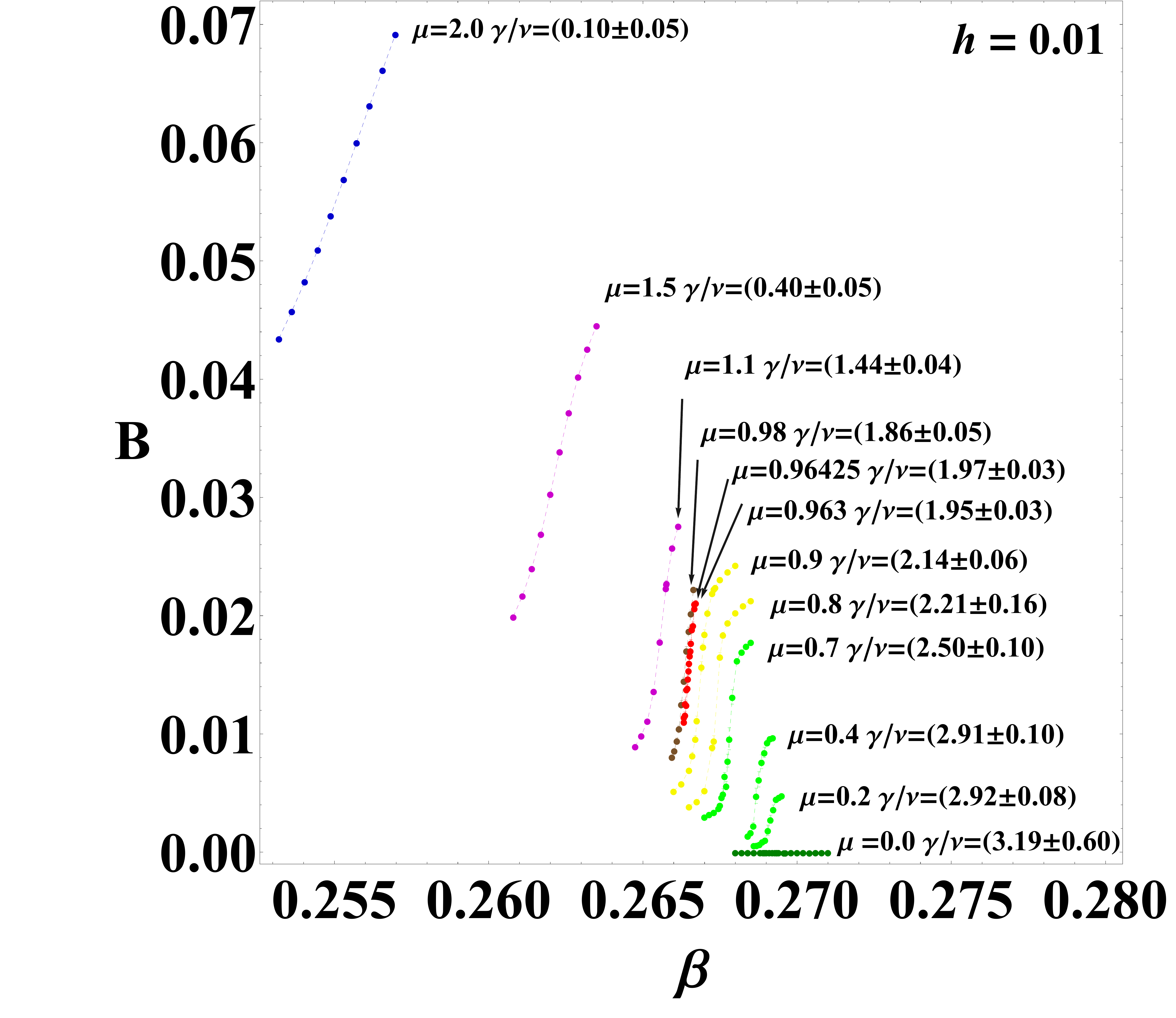}
\hfill
\includegraphics[width=0.48\textwidth]{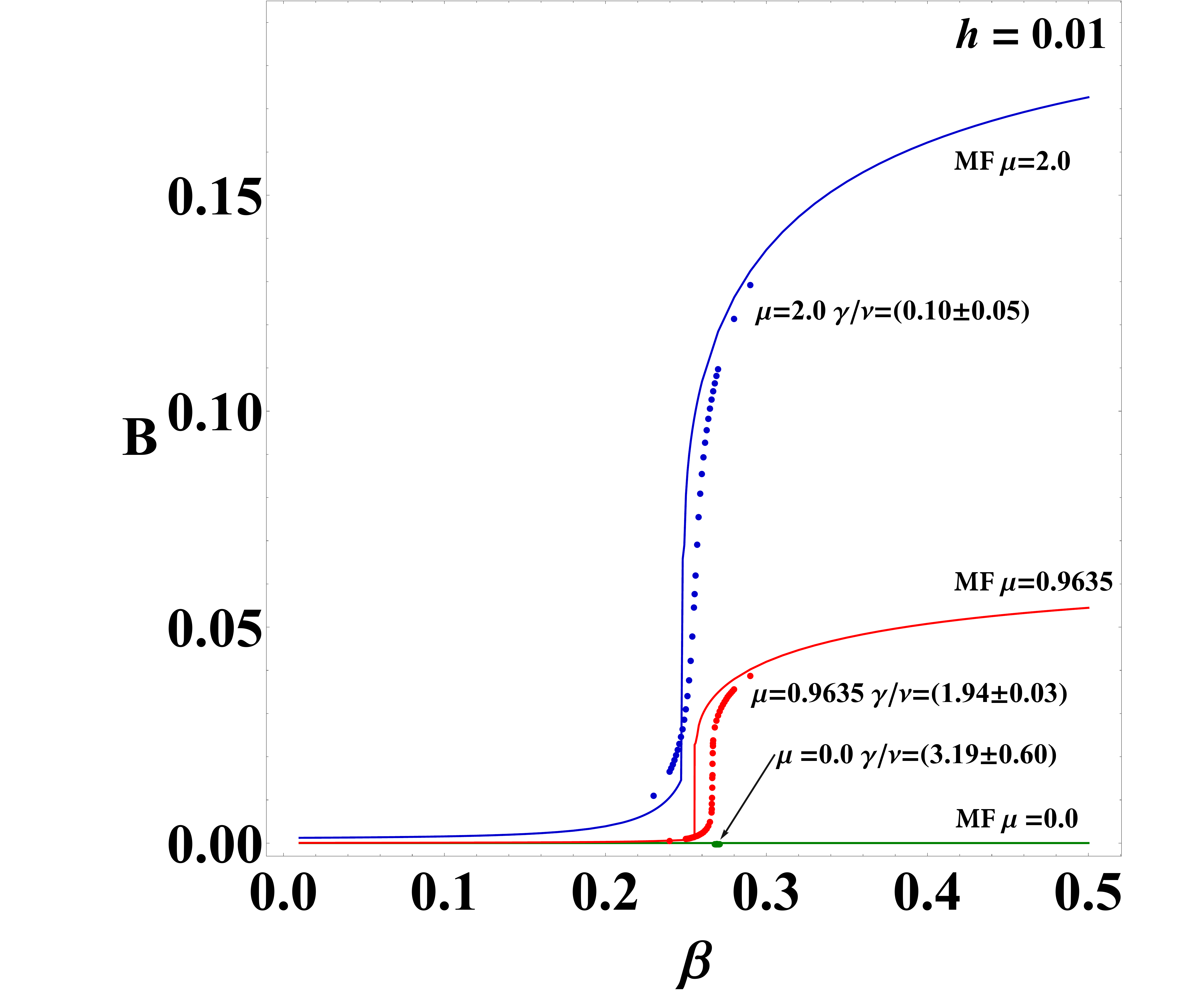}
\hfill
}
\caption{Behavior of the baryon density versus $\beta$ for fixed $h=0.01$ and
  for different values of $\mu$ on a 16$^3$ lattice.
  Left panel: results of simulations in the vicinity of a phase transition.
  Right panel: comparison of mean-field analysis and numerical results in three
  different phase regimes.}
\label{fig:2D_Baryon_Density_h0.01}
\end{figure}

The behavior of baryon density and quark condensate as functions of $\beta$
depends strongly on the phase of the system. Left panels of
Figs.~\ref{fig:2D_Quark_condensate_h0.01} and~\ref{fig:2D_Baryon_Density_h0.01} 
show the typical behavior of $Q$ and $B$ for $h=0.01$ and various values of the
chemical potential corresponding to different phases of the model. These phases 
are characterized as before by the values of $\gamma/\nu$ and can be
approximately read off from Fig.~\ref{fig:gamma_over_nu_vs_h_mu} or, more
precisely from the Table~\ref{table:beta_critic_and_gamma_over_nu_versus_h_nu}.
One observes a rapid change of both $Q$ and $B$ at the first order phase
transition. This rapid change becomes smoother and smoother when parameters
are gradually changed toward the second order line and then to the crossover
regime. The right panels of the same Figures compare Monte Carlo data with the
mean-field predictions in the three regions. One can conclude that 
mean-field reproduces numerical simulations with good accuracy.

\begin{figure}[htb]
  \centering
  {\hfill
    {\includegraphics[width=0.48\textwidth]{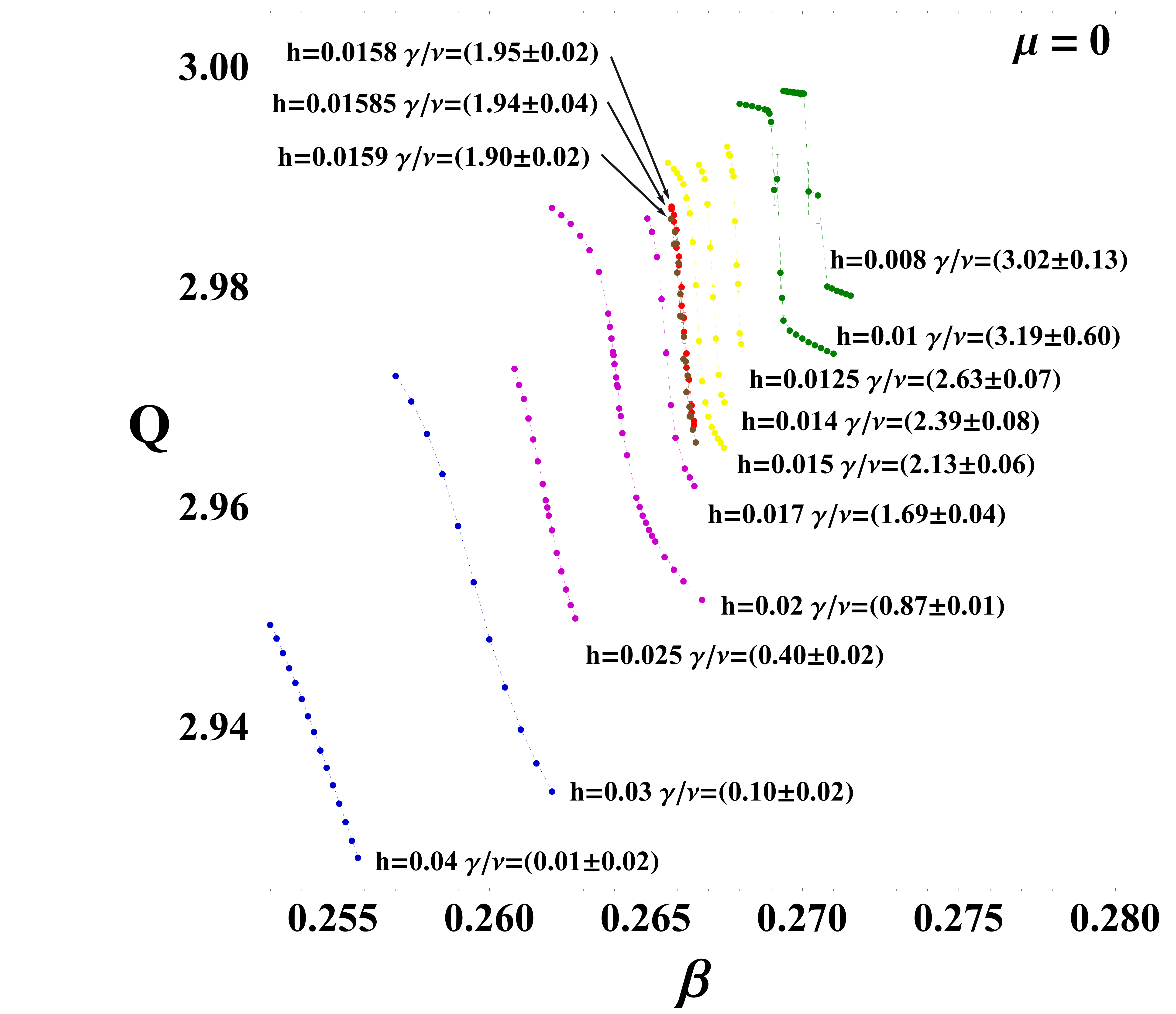}}
    \hfill
        {\includegraphics[width=0.48\textwidth]{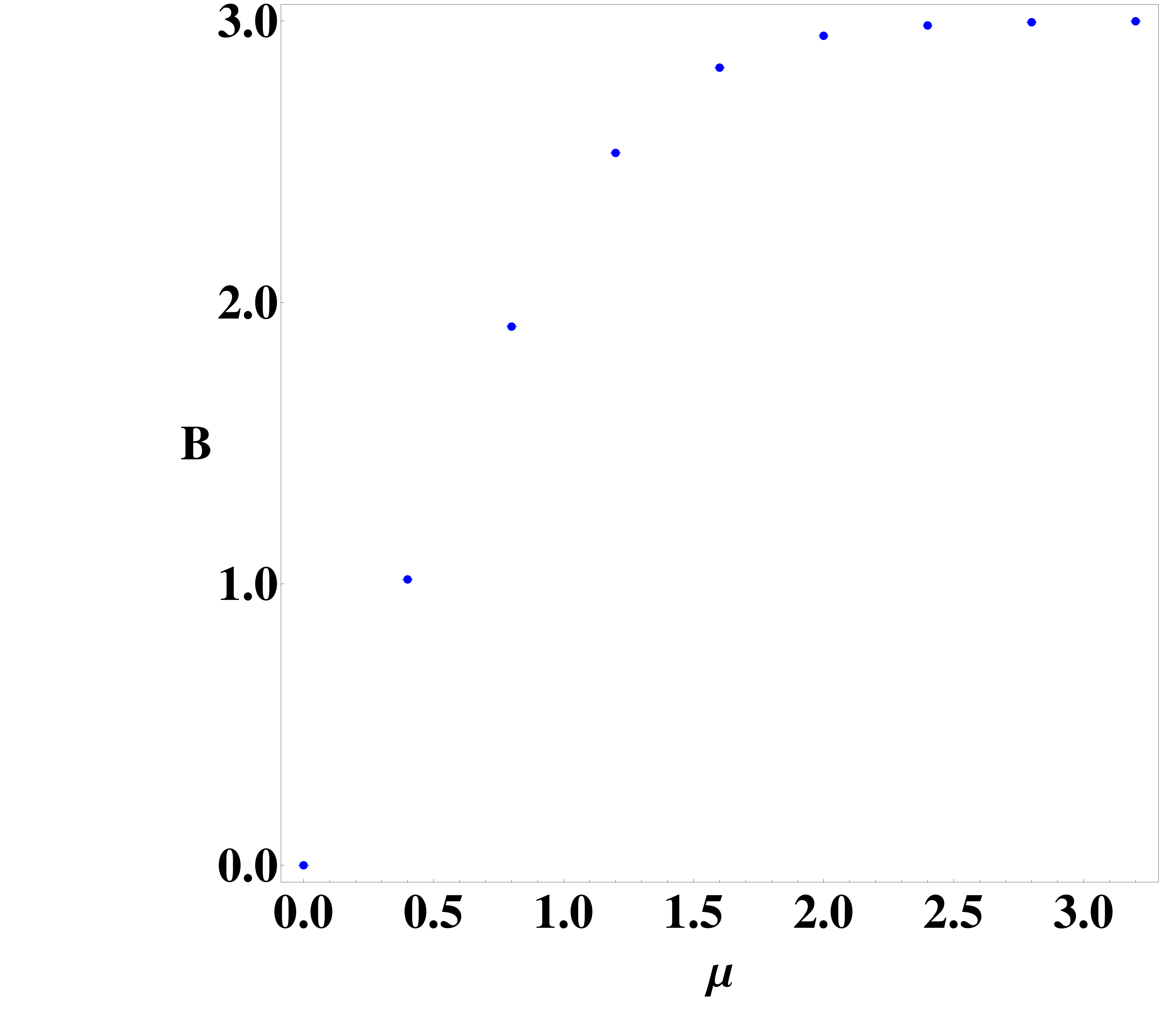}} \vspace{0.2cm}        
\hfill
   }     
  \caption{Left panel: behavior of the quark condensate versus $\beta$ for fixed
    $\mu=0$ and for different values of $h$ on a 16$^3$ lattice.
    Right panel: baryon density for $h=1$ and $\beta=0.1$ on a 16$^3$
    lattice.}
\label{fig:2D_Quark_condensate_mu_zero}
\end{figure}

The quark condensate as a function of $\beta$ is shown in
Fig.~\ref{fig:2D_Quark_condensate_mu_zero}(left) for vanishing value of the
chemical potential and several values of $h$. The right panel of the
same figure demonstrates the approach to the saturation of the baryon density
at zero quark mass, $h=1$.

As follows from Fig.~\ref{fig:comparison-high-h}, the baryon density $B$ is a
decreasing function of $\beta$ at sufficiently large $h=0.6$,
while Fig.~\ref{fig:2D_Baryon_Density_h0.01} shows that for small $h$ values
(large mass) $B$ is an increasing function of $\beta$. This conclusion is
supported by all three methods of calculations used in this paper. Therefore,
there should exists some value of the quark mass where the density changes its
qualitative behavior. We have not tried to estimate this value. 
  
\section{Large-distance behavior of the correlations} 

To see the impact of a non-zero chemical potential on the correlation function 
behavior, we calculated the two-point correlation functions for several values
of parameters. 
We considered six kinds of the correlation functions:

\begin{align}
 \Gamma_{nn}(r) &= \left\langle \Tr U(0) \Tr U(r) \right\rangle \ ,& 
 \Gamma_{rr}(r) &= \left\langle \Re \Tr U(0) \Re \Tr U(r) \right\rangle \ ,
 \nonumber \\
 \Gamma_{na}(r) &= \left\langle \Tr U(0) \Tr U^\dagger(r) \right\rangle \ ,&
 \Gamma_{ri}(r) &= \left\langle \Re \Tr U(0) \Im \Tr U(r) \right\rangle \ ,
 \nonumber \\
 \Gamma_{aa}(r) &= \left\langle \Tr U^\dagger(0) \Tr U^\dagger(r) \right\rangle \ ,&
 \Gamma_{ii}(r) &= \left\langle \Im \Tr U(0) \Im \Tr U(r) \right\rangle \ .
\end{align}

In the dual formulation the correlation functions can be written as
\begin{align}
 \Gamma_{nn}(r) &= \left\langle \frac{R_3(n(0) + 1, p(0))}{R_3(n(0), p(0))} 
				\frac{R_3(n(r) + 1, p(r))}{R_3(n(r), p(r))} \right\rangle
 \nonumber \\
 \Gamma_{na}(r) &= \left\langle \frac{R_3(n(0) + 1, p(0))}{R_3(n(0), p(0))} 
				\frac{R_3(n(r), p(r) + 1)}{R_3(n(r), p(r))} \right\rangle 
 \nonumber \\
 \Gamma_{aa}(r) &= \left\langle \frac{R_3(n(0), p(0) + 1)}{R_3(n(0), p(0))} 
				\frac{R_3(n(r), p(r) + 1)}{R_3(n(r), p(r))} \right\rangle \ .
 \label{gamma_dual}
\end{align}
These formulas work for $r > 0$. For $r = 0$ both shifts to the $n$, $p$
variables happen at one point, so only one ratio remains. The correlations
$\Gamma_{rr}$, $\Gamma_{ri}$ and $\Gamma_{ii}$ can be obtained as linear
combinations of $\Gamma_{nn}$, $\Gamma_{na}$ and $\Gamma_{aa}$.

The expressions~(\ref{gamma_dual}) become unusable when $h = 0$, and can 
have a bad convergence properties for very small $h$, or very large
$\mu$ values. 
We have checked by comparing the numerical results with the strong coupling
expressions for small $\beta$ values, that the results can be relied on for
$h > 0.005$ and $\mu < 3$. 

Since we work at non-zero $h$, the average traces can become non-zero,
introducing a constant term into the correlation function even in the
disordered phase. Due to that, we
introduce the subtracted correlation functions, subtracting the corresponding
average trace inside the correlations:
\begin{align}
 \Gamma_{nn, \rm sub}(r) &= \Gamma_{nn}(r) - \left\langle \Tr U \right\rangle^2 \ ,
 \nonumber \\
 \Gamma_{na, \rm sub}(r) &= \Gamma_{na}(r) - \left\langle \Tr U \right\rangle \left\langle \Tr U^\dagger \right\rangle \ ,
 \nonumber \\
 \Gamma_{aa, \rm sub}(r) &= \Gamma_{aa}(r) - \left\langle \Tr U^\dagger \right\rangle^2 \ .
\end{align}
For these subtracted correlations we expect an exponential decay,
\begin{equation}
 \Gamma(r) = A \frac{\exp(-m r)}{r} \;,
\label{corr-mass-gap}
\end{equation}
at least in the disordered phase.  

\begin{figure}[htb]
\centering{
  \includegraphics[width=0.75\textwidth]{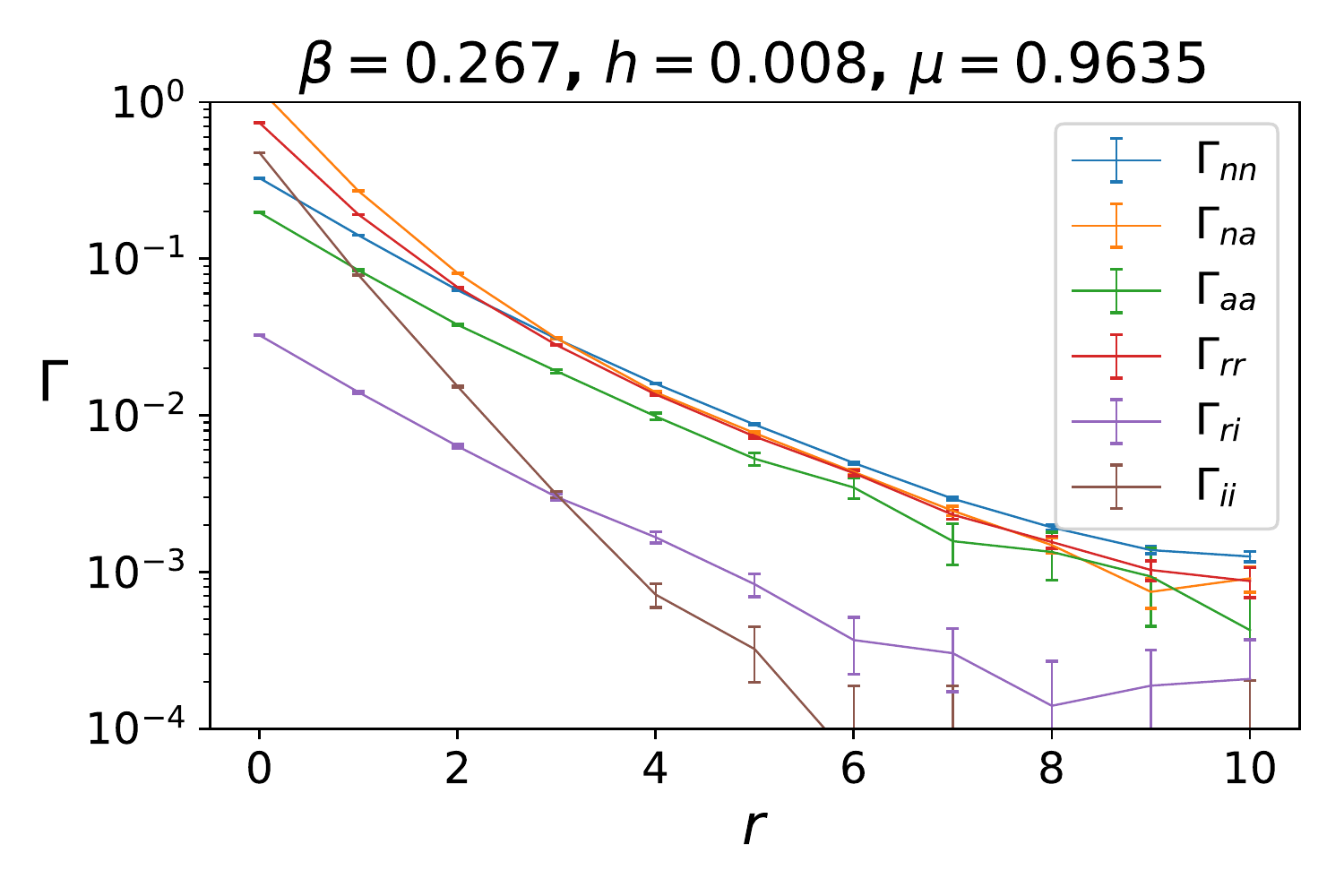}\\
  \includegraphics[width=0.75\textwidth]{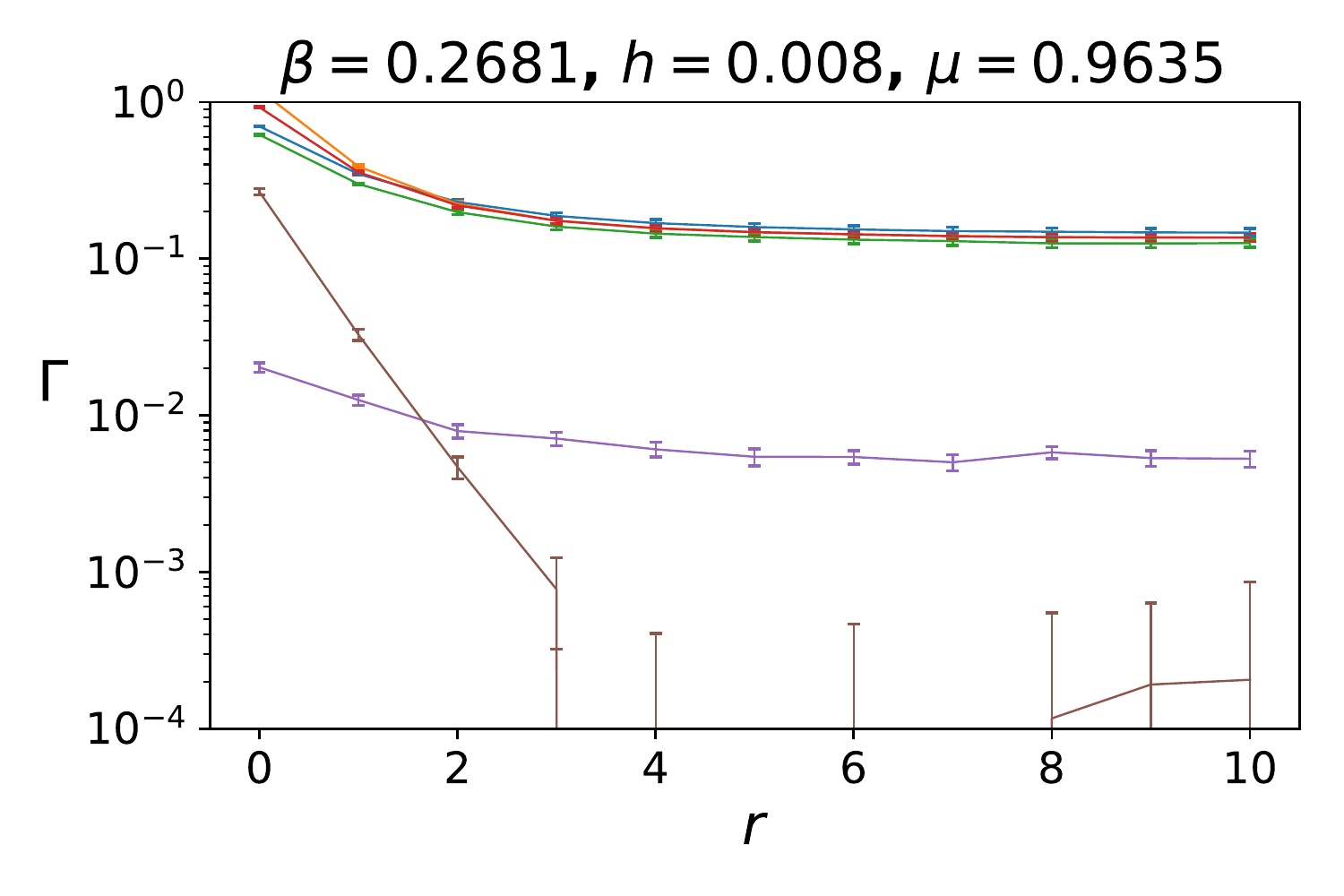} \\
  \includegraphics[width=0.75\textwidth]{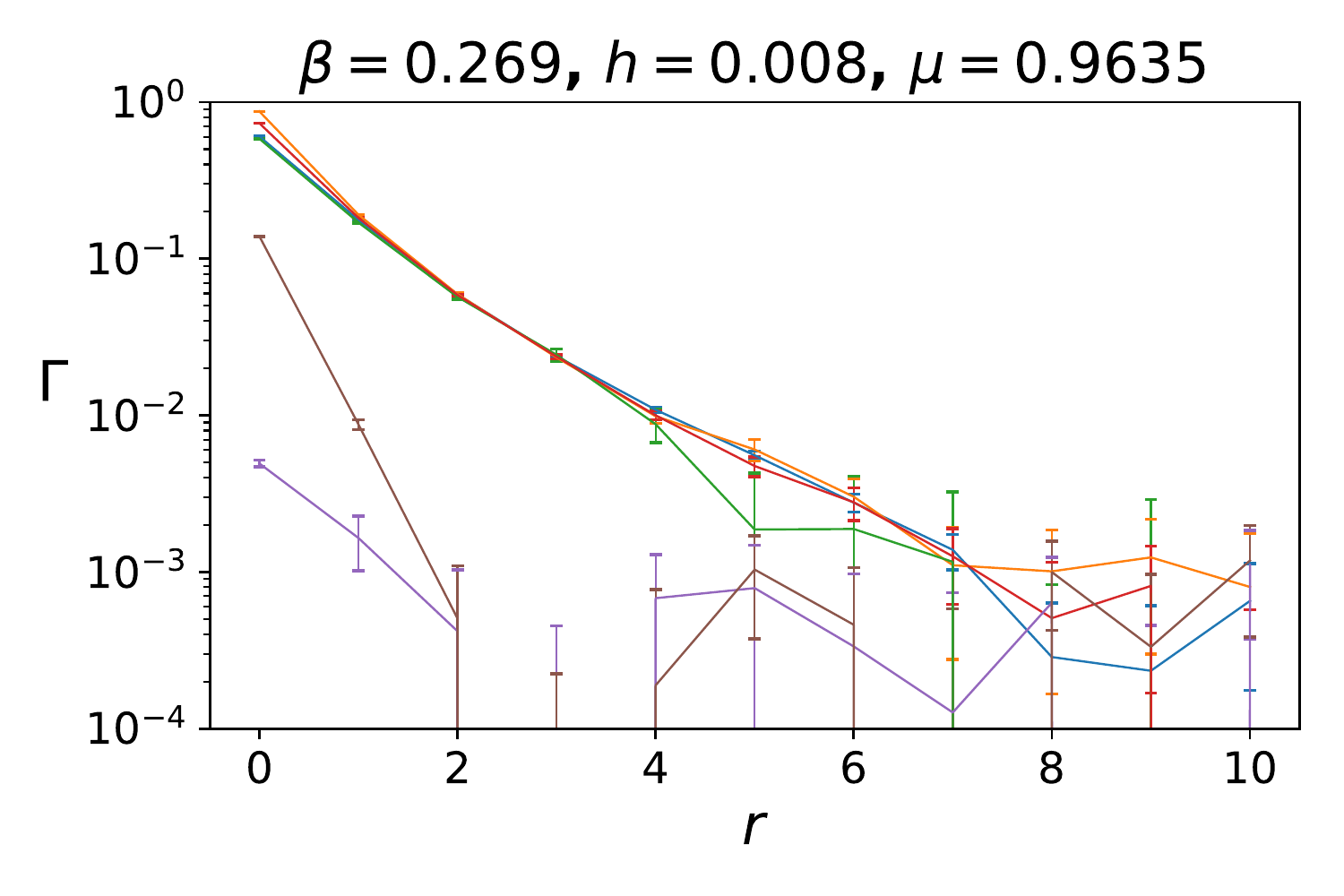} 
}
\caption{Behavior of the correlation functions on a $20^3$ lattice 
  for different values of parameter $\beta$ ($h=0.008$, $\mu=0.9635$).
  Top: disordered phase. Middle: near the first order phase transition point.
  Bottom: ordered phase. This legend applies also to the next figures.}
\label{fig:corr-first-order}
\end{figure}

\begin{figure}[htb]
\centering{
  \includegraphics[width=0.75\textwidth]{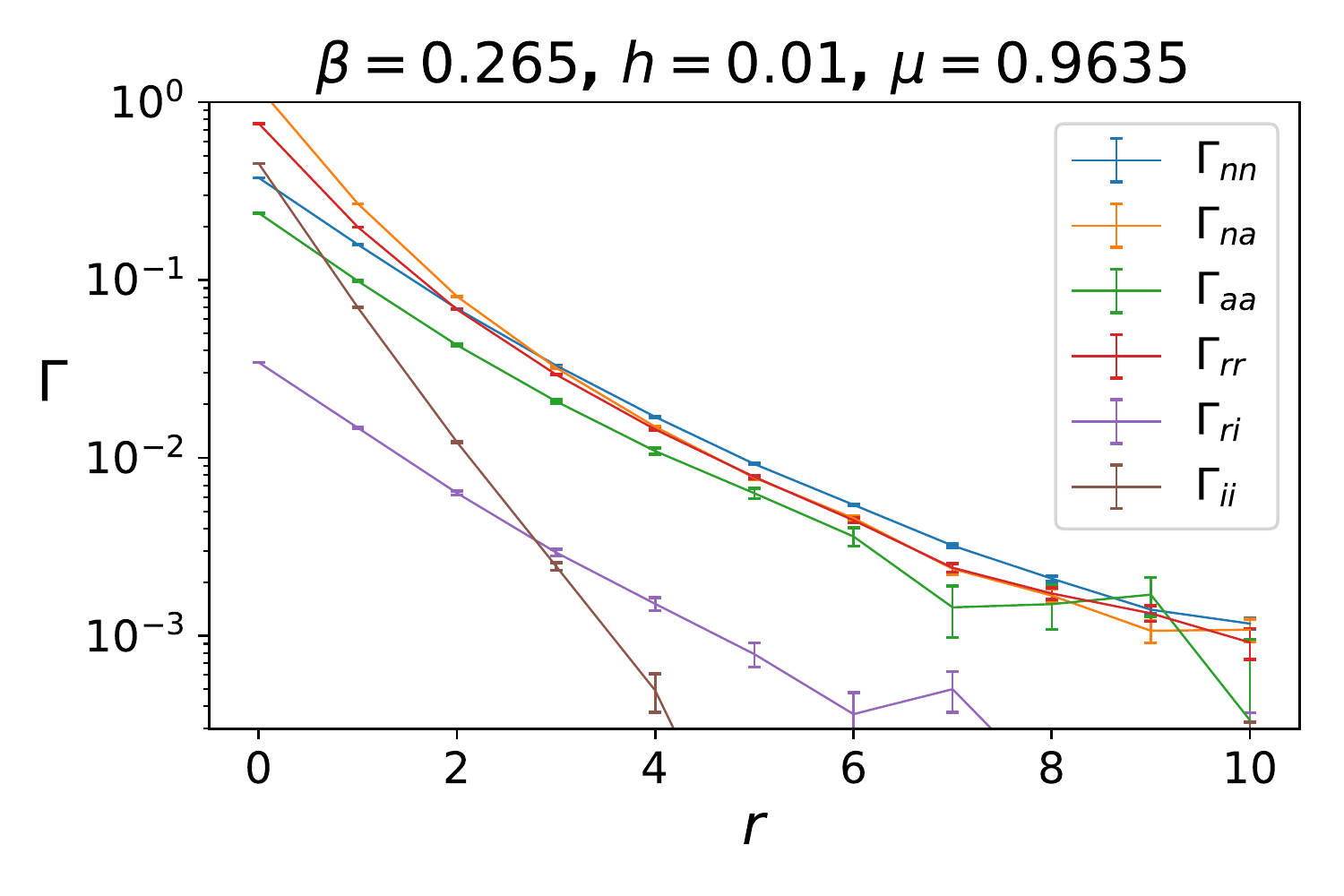} \\
  \includegraphics[width=0.75\textwidth]{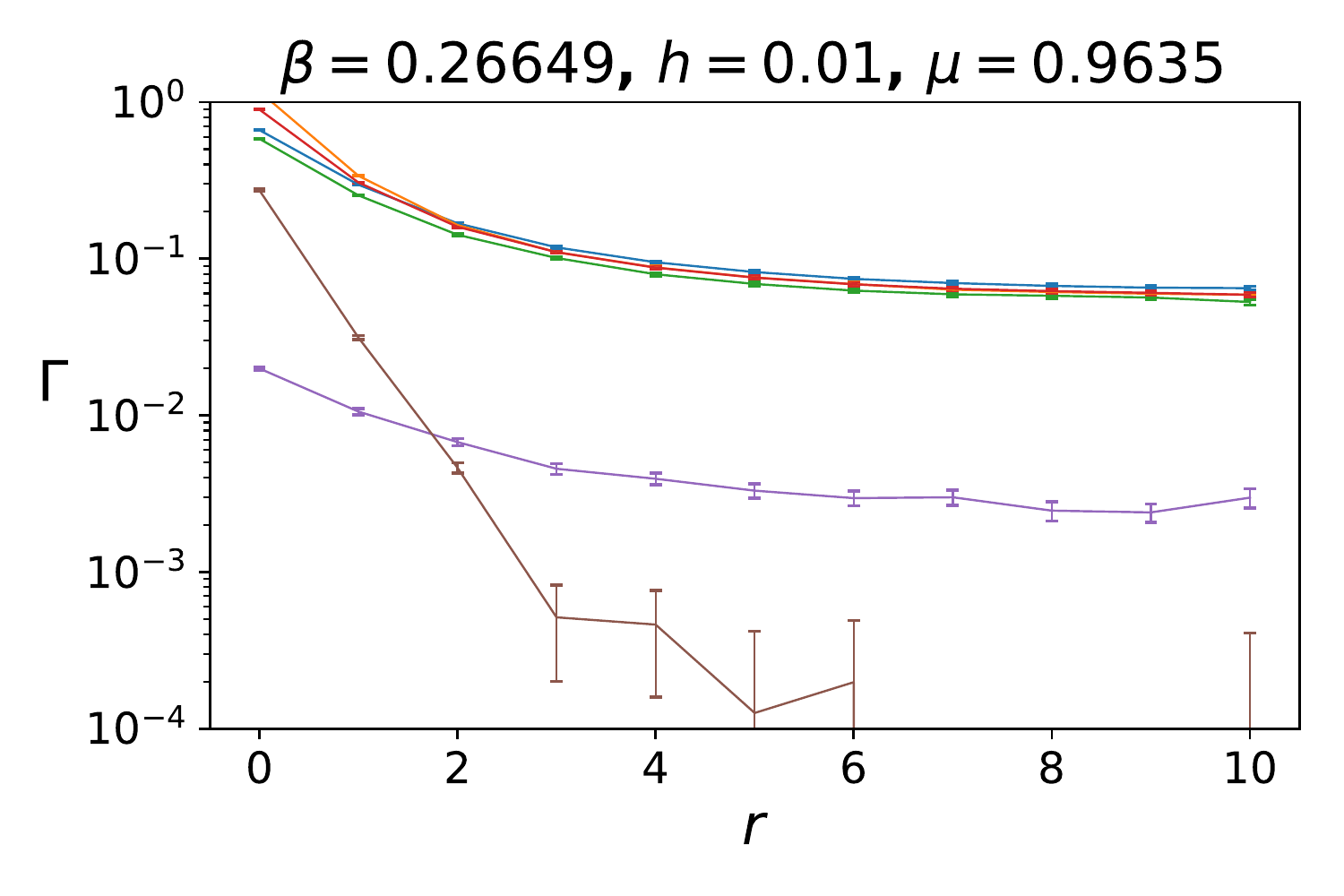} \\
  \includegraphics[width=0.75\textwidth]{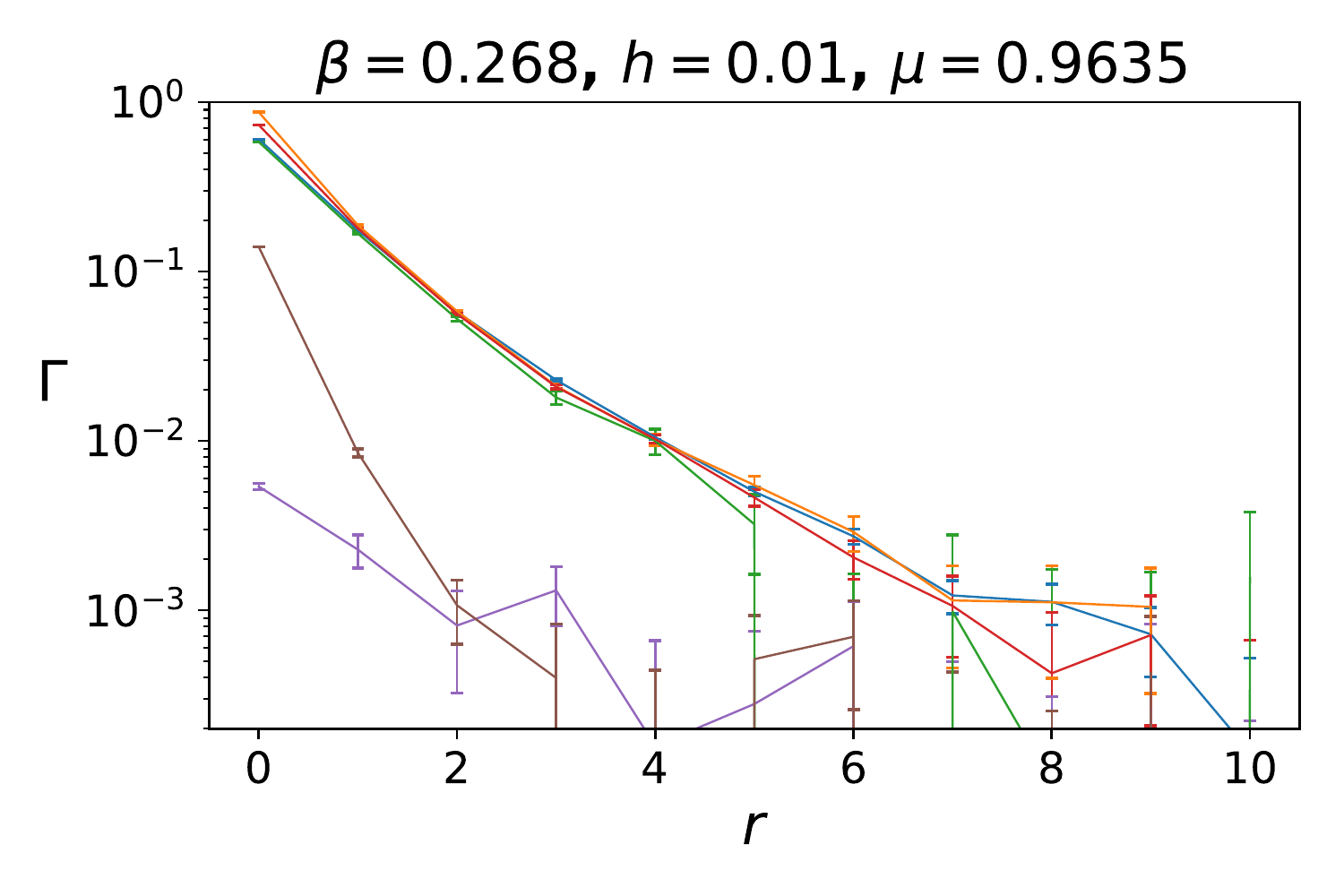} 
}
\caption{The same as Figure~\ref{fig:corr-first-order} for 
$h=0.01$, $\mu=0.9635$ (region of the second order phase transition).}
\label{fig:corr-second-order}
\end{figure}

\begin{figure}[htb]
\centering{
  \includegraphics[width=0.75\textwidth]{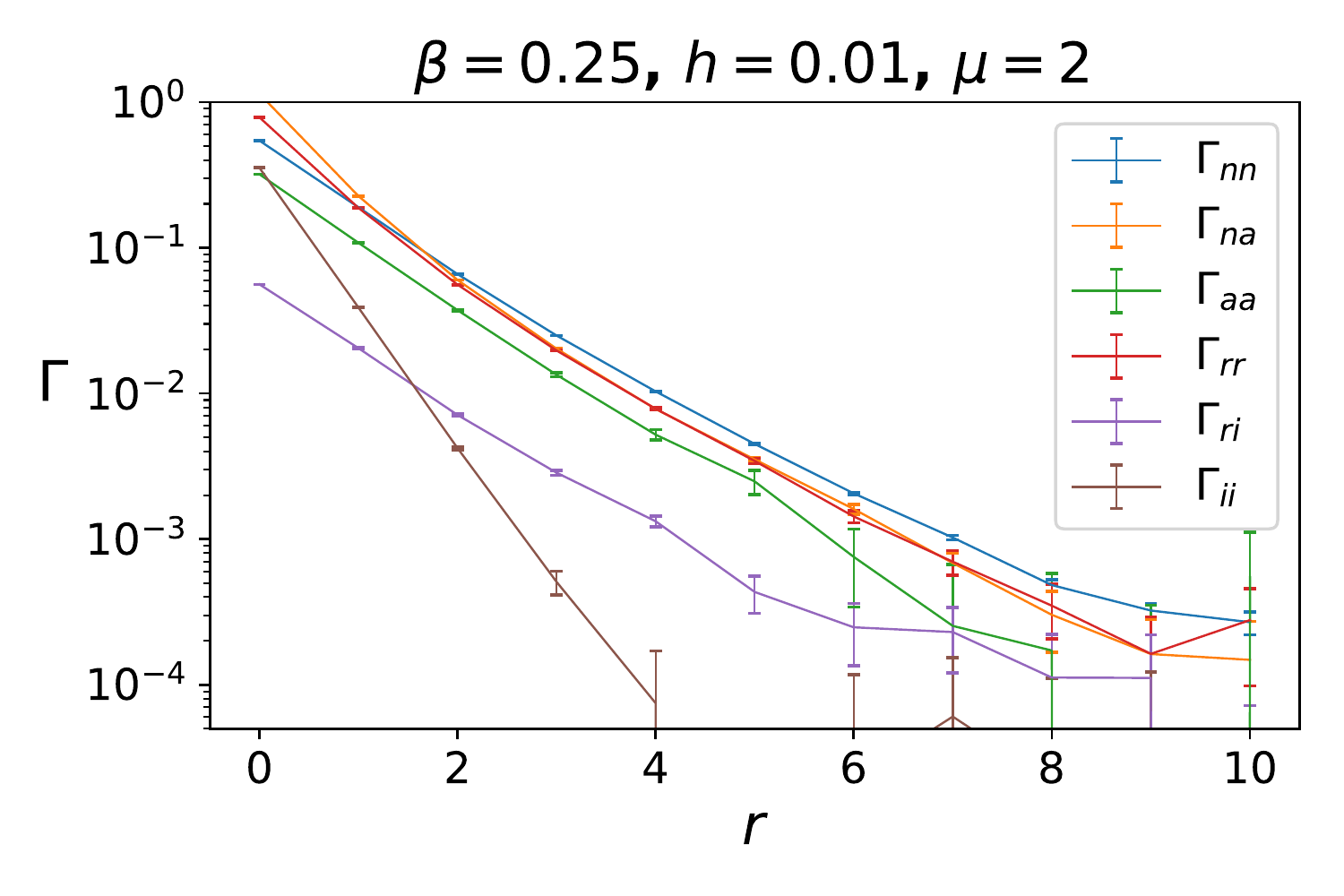} \\
  \includegraphics[width=0.75\textwidth]{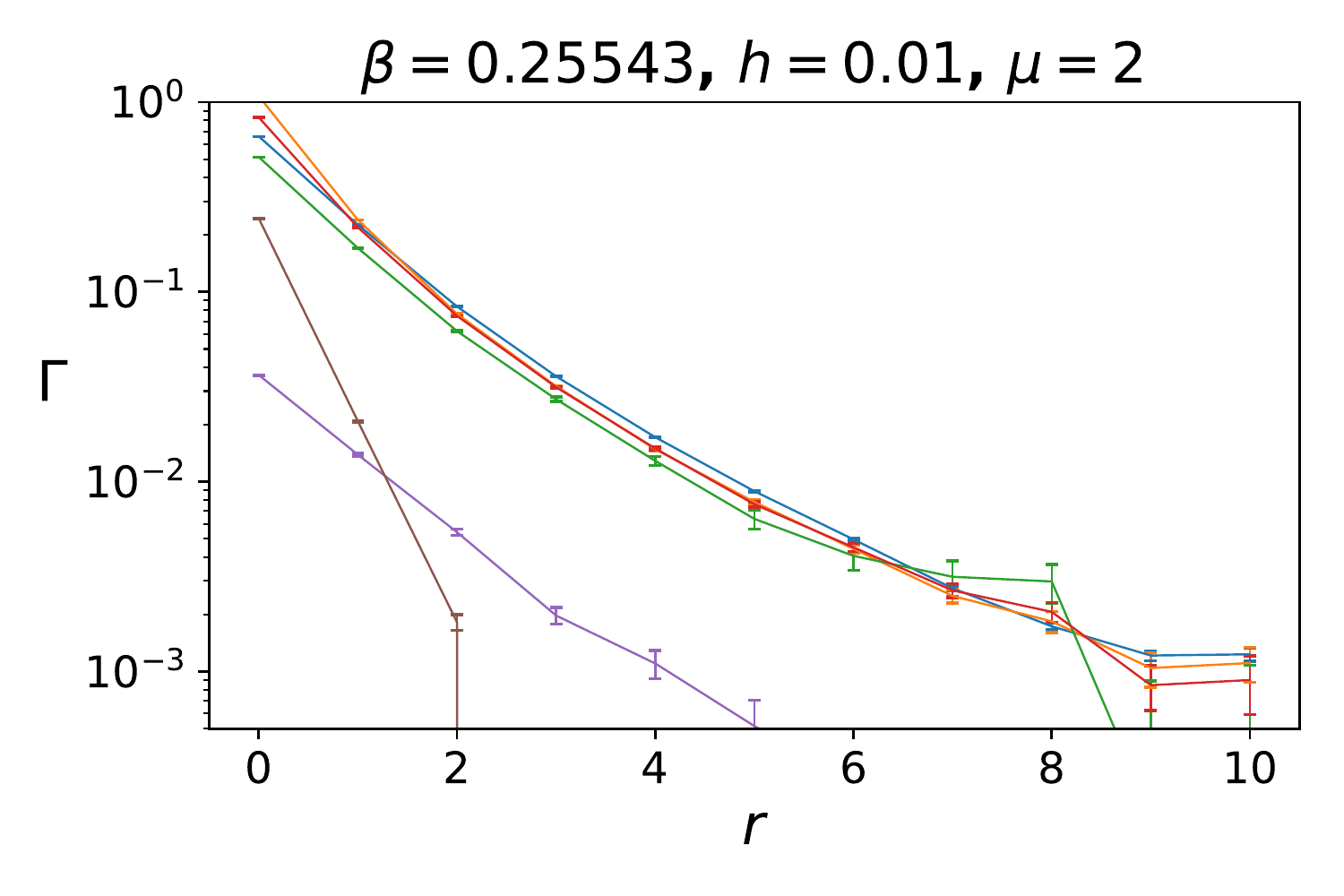} \\
  \includegraphics[width=0.75\textwidth]{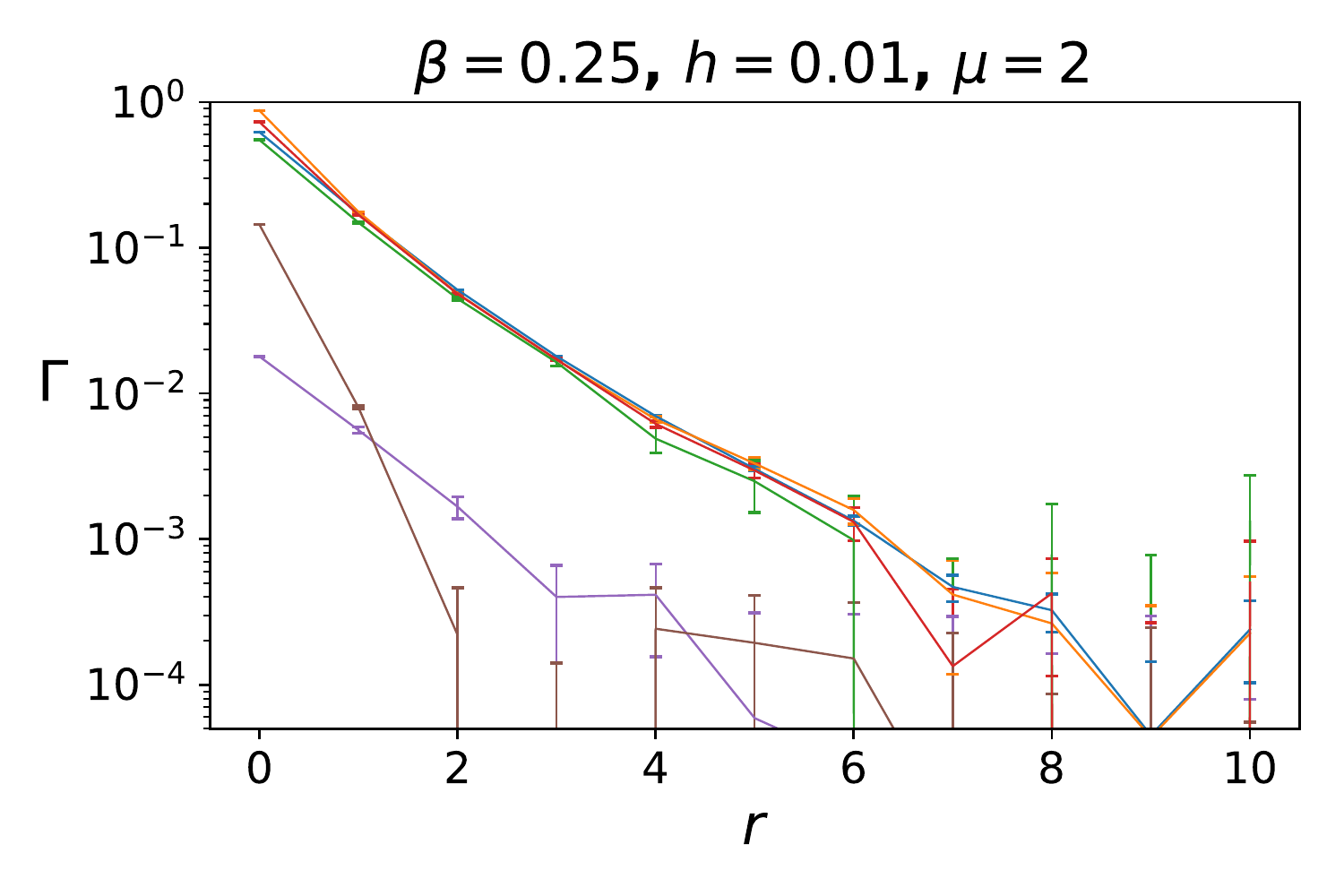} 
}
\caption{The same as Figure~\ref{fig:corr-first-order} for 
$h=0.01$, $\mu=2$ (crossover region).}
\label{fig:corr-crossover}
\end{figure}

Samples of correlation function behavior in different regions of the phase
diagram are shown in Figs.~\ref{fig:corr-first-order},
\ref{fig:corr-second-order} and~\ref{fig:corr-crossover}. 
One can see that, indeed, both in disordered and ordered phases,
the correlations decay exponentially. While the mass gap, 
corresponding to the slope of the plots, remains the same for $\Gamma_{nn}$,
$\Gamma_{na}$, $\Gamma_{aa}$, $\Gamma_{rr}$ and $\Gamma_{ri}$ correlation
functions, it is much larger for the $\Gamma_{ii}$ correlation function. 
Also the mass gap for the $\Gamma_{ii}$ correlation remains more or less
constant, and in particular does not vanish in the vicinity of the phase
transition. 

The difference in mass gaps can be explained by noting that at $\mu = 0$ 
$\Gamma_{rr}$ and $\Gamma_{ii}$ correspond to the color-magnetic 
and color-electric sectors having different mass gaps $m_M$ and $m_E$,
$m_M < m_E$ (see~\cite{bonati2018}).
While at non-zero $\mu$ these two sectors should mix, so all the correlators we
study should decay with $m_M$, it is possible that in our case the mixing is
small causing the real large distance mass gap for the color-electric sector
to be visible only on distances larger than the ones for which we have
reliable results. 

In the ordered phase we observe an increase of the correlation function slope 
(at the same $\beta$ values) with the increase of $\mu$, 
implying that the mass gap grows with $\mu$. This means an increase of the
screening effects: at finite density non-zero $\mu$ pushes system deeper in
the deconfined phase. 
This is in qualitative accordance with the results of~\cite{bonati2018}
obtained with imaginary $\mu$.

We will address these questions in more detail in a future work, which is under
preparation.
For now, we can note that at least in the disordered phase also the second
moment correlation length for the imaginary-imaginary correlations is
substantially different from the one for the real-real correlations.

\section{Summary}

Revealing the phase diagram of QCD at finite temperature and non-zero baryon
chemical potential, as well as the nature of strong interacting matter at high
temperatures, remains one of the important challenges of high energy physics.
In spite of enormous efforts, many aspects of these problems are still far
from unambiguous resolution. 
The several methods developed to solve the sign problem in finite density QCD
have their own advantages and drawbacks. Dual formulations of
  effective Polyakov loop models for heavy-dense QCD, as the one used
here, solve the sign problem completely, but are restricted so far to strong
coupled regions in the case of non-Abelian gauge theories. Even in this region
one can get valuable information about the phase structure of non-Abelian
models and behavior of various observables.   
The study presented here is in the same spirit of
Refs.~\cite{Gattringer12,Philipsen12,Delgado12}, though we have used a
different dual formulation of the Polyakov loop model, built in 
Ref.~\cite{Borisenko20}. To corroborate our findings we have also compared in
many cases simulation results with the strong coupling expansion of the dual
model and with the mean-field analysis. Let us briefly recapitulate our main
results.

\begin{itemize}
\item 
  The phase diagram of the model was studied in great details. We have
  classified three regions in the parameter space of the model
  $(\beta, h, \mu)$ according to the type of the critical behavior: first
  or second order phase transition, or crossover. 
  The values of the ratio of critical indices $\gamma/\nu$ is different in
  different regimes.  

\item 
  As main observables we computed expectation values of the Polyakov loop and
  its conjugate, the baryon density and the quark condensate.  

\item 
  Our dual formulation allows us to compute correlation functions of the
  Polyakov loops. We have presented some preliminary results for such
  correlations at non-zero chemical potential.  

\item 
  It is interesting to note that the mean-field results agree very well with
  numerical simulations both at zero and non-zero $\mu$. Also, the mean-field
  results are in good qualitative agreement with a similar analysis in
  Ref.~\cite{Greensite12}. 

\end{itemize}

The overall qualitative picture of the phase diagram and the behavior of all
observables fully agree with the picture described in
Refs.~\cite{Philipsen12,Delgado12}. 

All observables considered in this work have shown sensitivity to the
chemical potential. The general trend is that when $\mu$ is increased,
they exhibit a less steep variation across transition when the
coupling $\beta$ (which corresponds to the temperature in the underlying
QCD theory) is increased. Qualitatively, one can say that increasing $\mu$
plays effectively the same role as a reduction of the quark mass.

The most important direction for the future work is a detailed study of the
different correlations of the Polyakov loops and the extraction of screening
(electric and magnetic) masses at finite chemical potential. Also, an
investigation of the oscillating phase and the related complex masses can be
accomplished within our dual formulation. All these problems will be addressed
in a companion paper which is currently under preparation. 

\vspace{0.5cm}

{\bf \large Acknowledgements}

\vspace{0.2cm}

Numerical simulations have been performed on the ReCaS Data Center of
INFN-Cosenza.
O.~Borisenko acknowledges support from the National 
Academy of Sciences of Ukraine in frames of priority project 
"Fundamental properties of matter in the relativistic collisions 
of nuclei and in the early Universe" (No. 0120U100935).
The author V.~Chelnokov acknowledges support by the Deutsche
Forschungsgemeinschaft (DFG, German Research Foundation) through the
CRC-TR 211 'Strong-interaction matter under extreme conditions' --
project number 315477589 -- TRR 211.
E.~Mendicelli was supported in part by a York University Graduate
Fellowship Doctoral - International.
A.~Papa acknowledges support from the Istituto Nazionale di Fisica Nucleare
(INFN) through the NPQCD project.

\vspace{0.5cm}


\end{document}